\documentclass[review]{elsarticle}

\usepackage{wrapfig}

\usepackage[utf8]{inputenc} 
\usepackage[T1]{fontenc}    
\usepackage{hyperref}       
\usepackage{url}            
\usepackage{booktabs}       
\usepackage{nicefrac}       
\usepackage{microtype}      

\usepackage{times}
\usepackage{helvet}
\usepackage{courier}

\usepackage{tikz}
\usepackage{amsmath,amsfonts, amssymb}       

\usetikzlibrary{arrows.meta,
                calc, chains,
                quotes,
                positioning,
                shapes.geometric,bayesnet}

\def\mmu{\boldsymbol{\mu}}

\def\IND{\perp\!\!\!\perp}
\def\I{\mathcal I}

\def\K{\mathcal K}

\def\z{\mathbf{z}}
\def\Z{\mathbf{Z}}

\def\real{\mathrm{I\!R}}

\def\y{\mathbf y}

\def\x{\mathbf x}

\def\eps{\varepsilon}

\def\I{\mathbf I}
\def\one{\mathbf 1}

\def\E{\mathbb E}
\def\V{\mathbb V}

\def\N{{\mathcal N}}

\def\M{{\mathcal M}}
\def\K{\mathbf K}
\def\H{\mathbf H}
\def\L{\mathbf L}
\def\Ell{\mathcal{L}}
\def\Jll{\mathcal{J}}
\def\PPhi{\mathbf \Phi}
\def\pphi{\boldsymbol \phi}

\def\t{^\top}
\def\ttheta{\boldsymbol \theta}
\def\ppi{\boldsymbol \pi}

\def\HSIC{\mathrm{HSIC}}

\DeclareMathOperator*{\argmin}{arg\,min}
\DeclareMathOperator*{\argmax}{arg\,max}

\DeclareMathOperator{\tr}{tr}

\newtheorem{thrm}{Theorem}

\usepackage{lineno,hyperref}
\modulolinenumbers[5]

\journal{Journal of Machine Learning with Applications}

\biboptions{authoryear}

\begin{document}

\begin{frontmatter}

\title{A Causal Direction Test for Heterogeneous Populations}


\author[noahmont,poly]{Vahid Partovi Nia}
\author[noahmont]{Xinlin Li}
\author[mcgilluni]{Masoud Asgharian}
\author[chineseuni]{Shoubo Hu}
\author[noahhong]{Yanhui Geng}
\author[noahhong]{Zhitang Chen}


\address[noahmont]{Huawei Noah's Ark Lab, Suit 201, 7101 Park avenue, Montreal, Quebec H3N 1X9, Canada. \texttt{vahid.partovinia@huawei.com, xinlinli1@huawei.com}}
\address[noahhong]{Huawei Noah's Ark Lab, Units 525-530, Core Building 2,
Hong Kong Science Park, Shatin, Hong Kong. \texttt{chenzhitang2@huawei.com}, \texttt{geng.yamhui@huawei.com}}
\address[mcgilluni]{McGill University, Department of Mathematics and Statistics, 805 Rue Sherbrooke West, Montreal, QC H3A 2K6, Canada  \texttt{masoud.asgharian2@mcgill.ca}}
\address[chineseuni]{The Chinese University of Hong Kong, Department of Computer Science and Engineering, Ho Sin-Hang Engineering Building,
The Chinese University of Hong Kong,
Shatin N.T., Hong Kong. \texttt{sbhu@cse.cuhk.edu.hk}. }
\begin{abstract}
A probabilistic expert system  emulates the decision-making ability of a human expert through a directional graphical model. The first step in building such systems is to understand data generation mechanism. To this end, one may try to 
decompose a multivariate distribution into product of several conditionals, and evolving a blackbox machine learning predictive models towards transparent  cause-and-effect discovery. Most causal models assume a single homogeneous population, an assumption that may fail to hold in many applications. We show that when the  homogeneity assumption is violated, causal models developed based on such assumption can fail to identify the correct causal direction. We propose an adjustment to a commonly used causal direction test statistic by using a $k$-means type clustering algorithm where both the labels and the number of components are estimated from the collected data to adjust the test statistic. Our simulation result show that the  
proposed adjustment significantly improves the performance of the causal direction test statistic for heterogeneous data.  We study large sample behaviour of our proposed test statistic and demonstrate the application of the proposed method using real data.
\end{abstract}

\end{frontmatter}


\newpage
\section{Introduction}
\label{sect:intro}
Causal inference is one of the most fundamental concepts in learning. Teaching the machine how to find  \emph{cause-and-effect} relationship, often using a mathematical model, is as essential as teaching children how to connect the dots.
Causality is perhaps as old as human endeavour for learning and has historically developed along with the development of human knowledge in almost any domain of science. Scientific studies tend to draw cause and effect conclusions after observing associations. However, most observed associations translate to causal conclusions only under certain conditions. 
Early causal problem formulation based on data observation appears in statistics \citep{Neyman_Causal_1923}, economics \citep{Heckman_causal_1976}, medicine \citep{greenland1999causal}, and computer science \citep{pearl1986fusion} among others. Research on causality typically starts with hypothesizing a cause and setting up an experimental study in which association can be translated into causal relationship \citep{fisher1926arrangement}. In many applications, however, logistic or other constraints may preclude the possibility of conducting experimental studies. Methods have therefore been developed for cause-and-effect conclusions to be   drawn from data collected in observational studies. 

With the advent of technology, modern applications often include a large number of variables, many possibly spurious, measured on each subject under study. Identifying causal direction, especially in such high dimensional settings, albeit challenging, but crucial to gain insights into the data generating mechanism and hence data interpretation, and also a deeper understanding of data structure. 
There are recent efforts in using causal analysis to interpret large dimensional data in decision making \citep{wu2010linking}, in natural language processing \citep{dehkharghani2014sentimental}, in transportation \citep{kayikci2014causal}, and in genetics \citep{schadt2005integrative}, among others. 

\begin{figure}
    \centering
    \includegraphics[height=0.2\textheight]{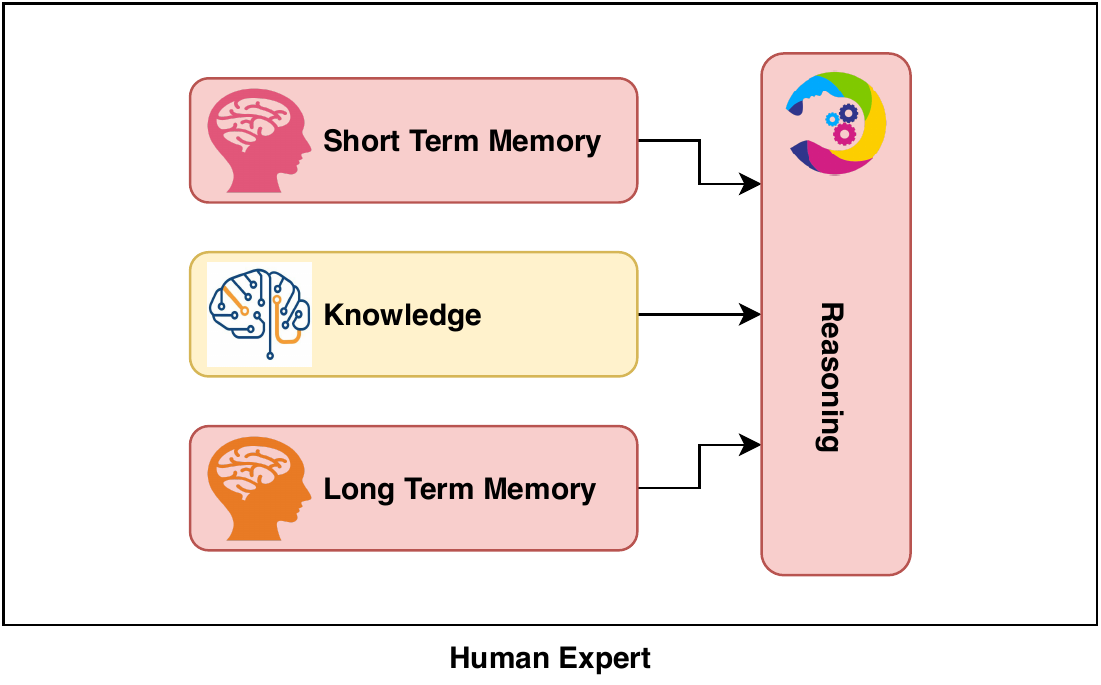}\\
    \bigskip
~~~~~~    \includegraphics[height=0.2\textheight]{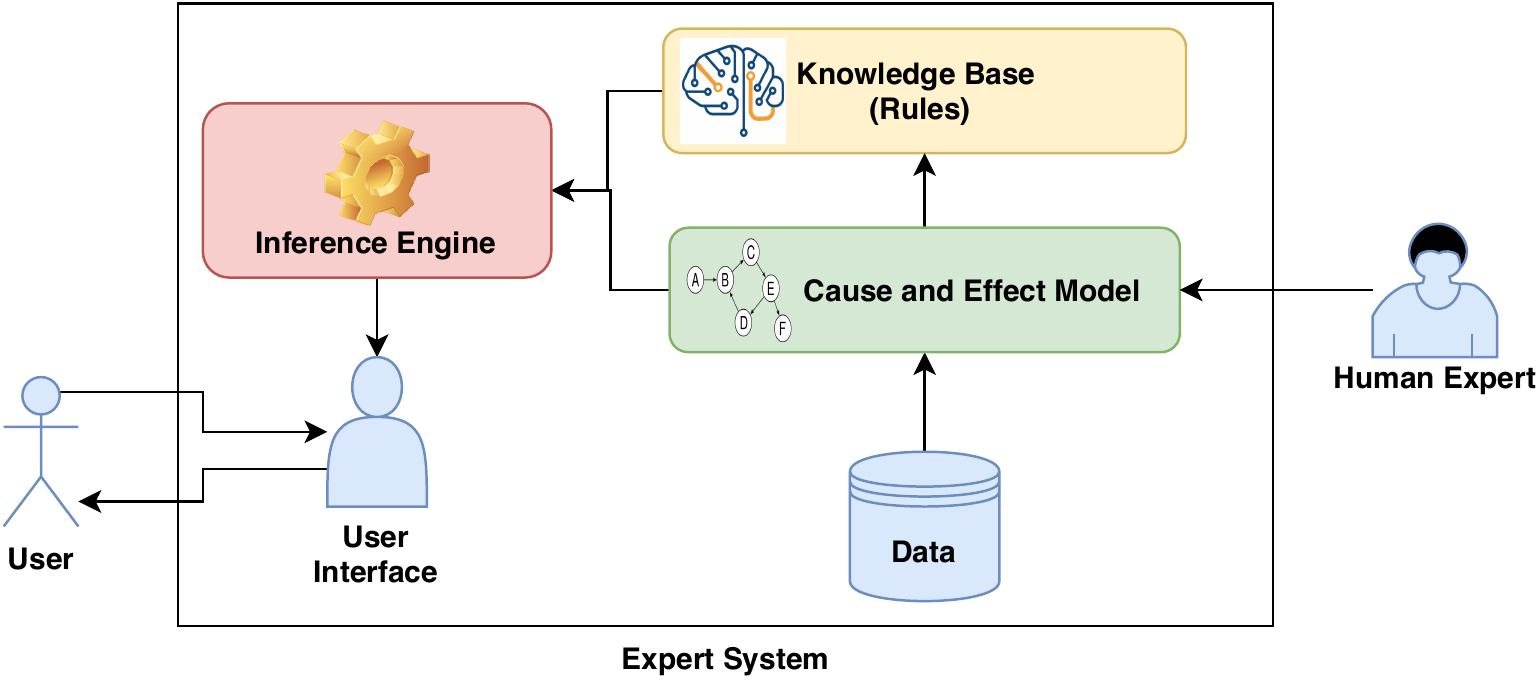}
    \caption{A human expert (top panel) versus an expert system embedded with automatic reasoning (bottom panel) in which cause and effect model attaches data to the knowledge base.}
    \label{fig:expertsystem}
\end{figure}

In many applications causal conclusion is drawn for  multiple variables, through a cause and effect model also known as belief network. Such models decompose a multivariate distribution into several conditional distributions. Many of such decompositions are theoretically equivalent and detecting which one is the actual data generating mechanism often requires domain knowledge of a human expert, see Figure~\ref{fig:expertsystem}. 



The precursor to establishing the cause and effect relationship between several variables is to identify the causal direction between a pair of variables. 
It transpires that inferring causal direction between a pair of variables, say $X$ and $Y$, should be closely tied to the conditional distributions of $X$ given $Y$ and of $Y$ given $X$.   
Assume $(X,Y)$ follows a complex joint distribution, say $p(x,y)$. Following the {\emph independent causal mechanisms principles} postulate  \citep{scholkopf2012causal}, causal direction inference looks for evidence in the observed data to prefer a certain conditional decomposition, either  $p(x,y) = p(y\mid x) p(x)$ or $p(x,y)= p(x\mid y) p(y)$. In the former decomposition $x$ causes $y$ or $x\to y$  and in the latter decomposition $y$ causes $x$ or $y\to x$. In theory, both decompositions are valid and therefore inferring a causal direction without further assumptions is ill-defined and \emph{unidentifiable}. Under more structural assumptions, such as the Additive Noise Model (ANM), the direction of decomposition becomes \emph{identifiable}, and observed data can be used to infer the cause and effect relationship.

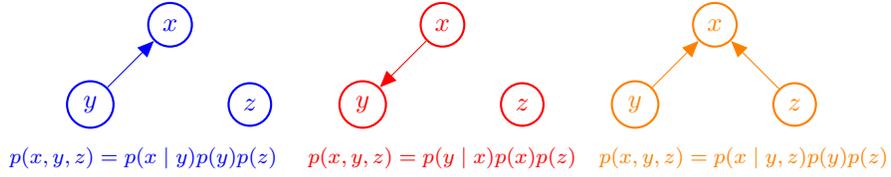
\begin{figure}
\begin{tikzpicture}[main/.style = {node distance={15mm}, thick, draw, circle}] 
\node[main, color=blue] (y1) {$y$}; 
\node[main, color=blue] (x1) [above right of=y1] {$x$};
\node[main, color=blue] (z1) [below right of=x1] {$z$};
\node[align=left]  (t1) [below right of =y1] {\footnotesize \color{blue} $p(x,y,z) = p(x\mid y) p(y)p(z)$};
\draw[->,color=blue] (y1) -- (x1);

\node[main, color=red] (y2) [right of=z1] {$y$};
\node[main, color=red] (x2) [above right of=y2] {$x$};
\node[main, color=red] (z2) [below right of=x2] {$z$};
\node[align=left]  (t2) [below right of =y2] {\footnotesize ~~~~~~~~~~\color{red}$p(x,y,z) = p(y\mid x) p(x)p(z)$};
\draw[->, color=red] (x2) -- (y2);

\node[main,color=orange] (y3) [right of=z2] {$y$};
\node[main,color=orange] (x3) [above right of=y3] {$x$};
\node[main,color=orange] (z3) [below right of=x3] {$z$};
\node[align=left]  (t3) [below right of =y3] {\footnotesize ~~~~~~~~~~~~~~~~~~~~~\color{orange}$p(x,y,z) = p(x\mid y,z) p(y)p(z)$};
\draw[->,color=orange] (y3) -- (x3);
\draw[->,color=orange] (z3) -- (x3);


\end{tikzpicture} 
\caption{Examples of simple cause and effect models for three variables.  When $y$ causes $x$, i.e. $p(x,y,z) = p(x\mid y) p(y) p(z)$, left panel. When $x$ causes $y$, i.e. $p(x,y,z) = p(y\mid x) p(x) p(z)$, middle panel. When $y$ and $z$ cause $x$, i.e. $p(x,y,z) = p(x\mid y, z) p(y) p(z)$, right panel. }
\label{fig:toydirection}
\end{figure}

ANM \citep{hoyer2009nonlinear} represents the effect as a function of the cause with an additive independent noise, i.e. $y = f(x) + \varepsilon$, in which $f$ is a nonlinear deterministic smooth function 
and $\eps\sim p(\eps)$ is an independent noise. There is no backward model of the form $x = g(y) + \eps$ that admits an ANM in the anti-causal direction unless the backward noise depend on  $y$. This shows a causal direction can be examined based on testing $\eps \IND x.$ One may first fit a smooth regression model to predict $\hat y = \hat f(x)$, predict the noise $\hat \eps= y-\hat y$, and then test if the residuals $\hat\eps$ is independent of the predictor $x$, $\hat \eps \IND x$ \citep{hoyer2009nonlinear}. This test of independence can be used as an evidence for inferring the causal direction for a pair of variables empirically.

Most causal inference approaches assume a single causal model for the observed data \citep{shimizu2006linear,zhang2009identifiability,janzing2010causal} so they are suitable for homogeneous data. In many applications, however, data are collected from several different sources. Due to the unknown data generation process and variability in the data source, sampling scheme, sampling conditions, etc there is no guarantee on the viability of such homogeneity assumptions in practice. Naive use of the existing clustering algorithms misleads causal direction inference. Moreover, each sub cluster may declare its own causal direction to confuse the ultimate  judgment.

When observations are generated from a non-homogeneous population that is comprised of several homogeneous sub-populations, the number of homogeneous sub-populations affects the causal direction test statistic performance to a great extent. The distribution of the causal direction test statistic needs to be adjusted when the homogeneity assumption fails to hold.  

\begin{figure}
\centering
		\includegraphics[width=0.45\linewidth]{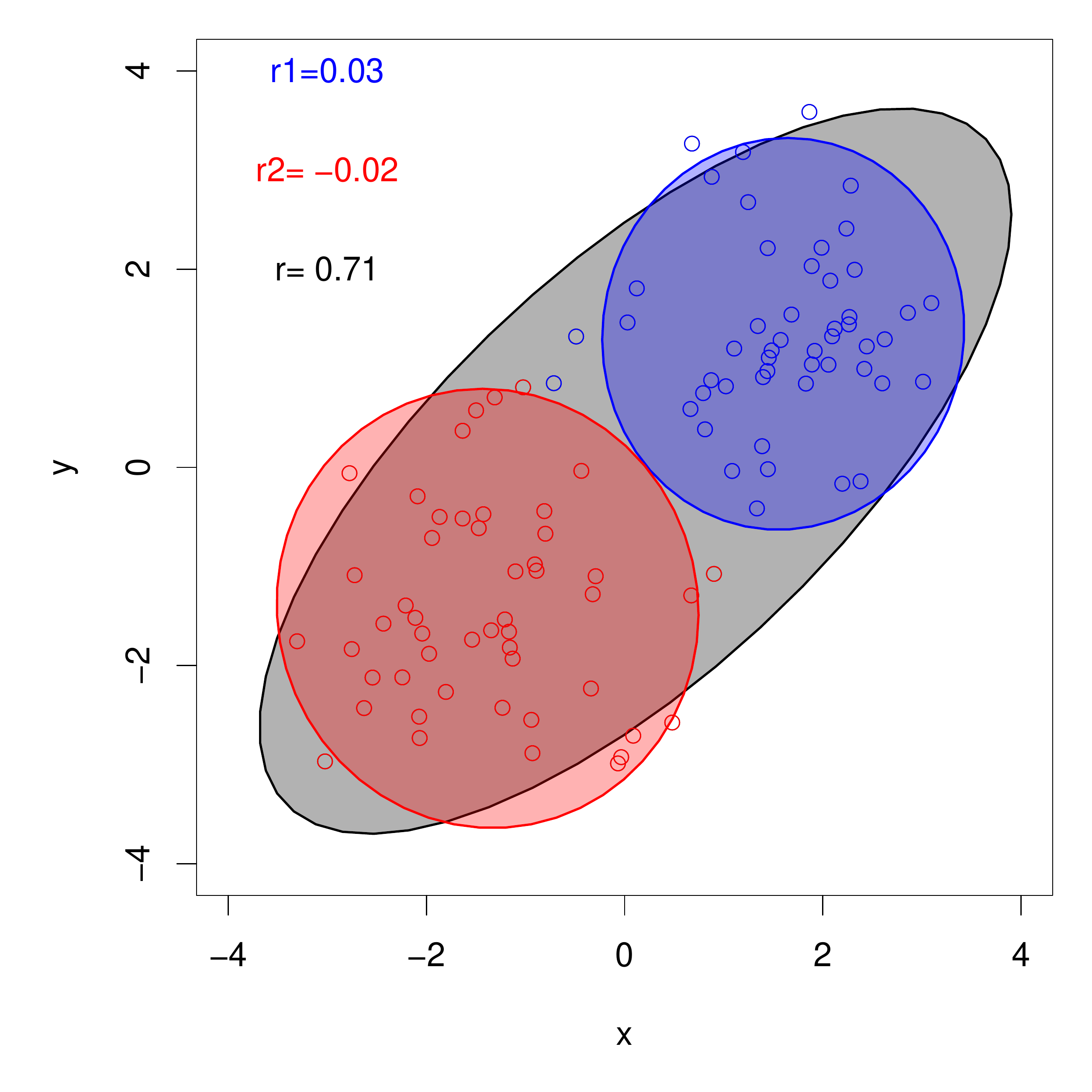}
		\includegraphics[width=0.45\linewidth]{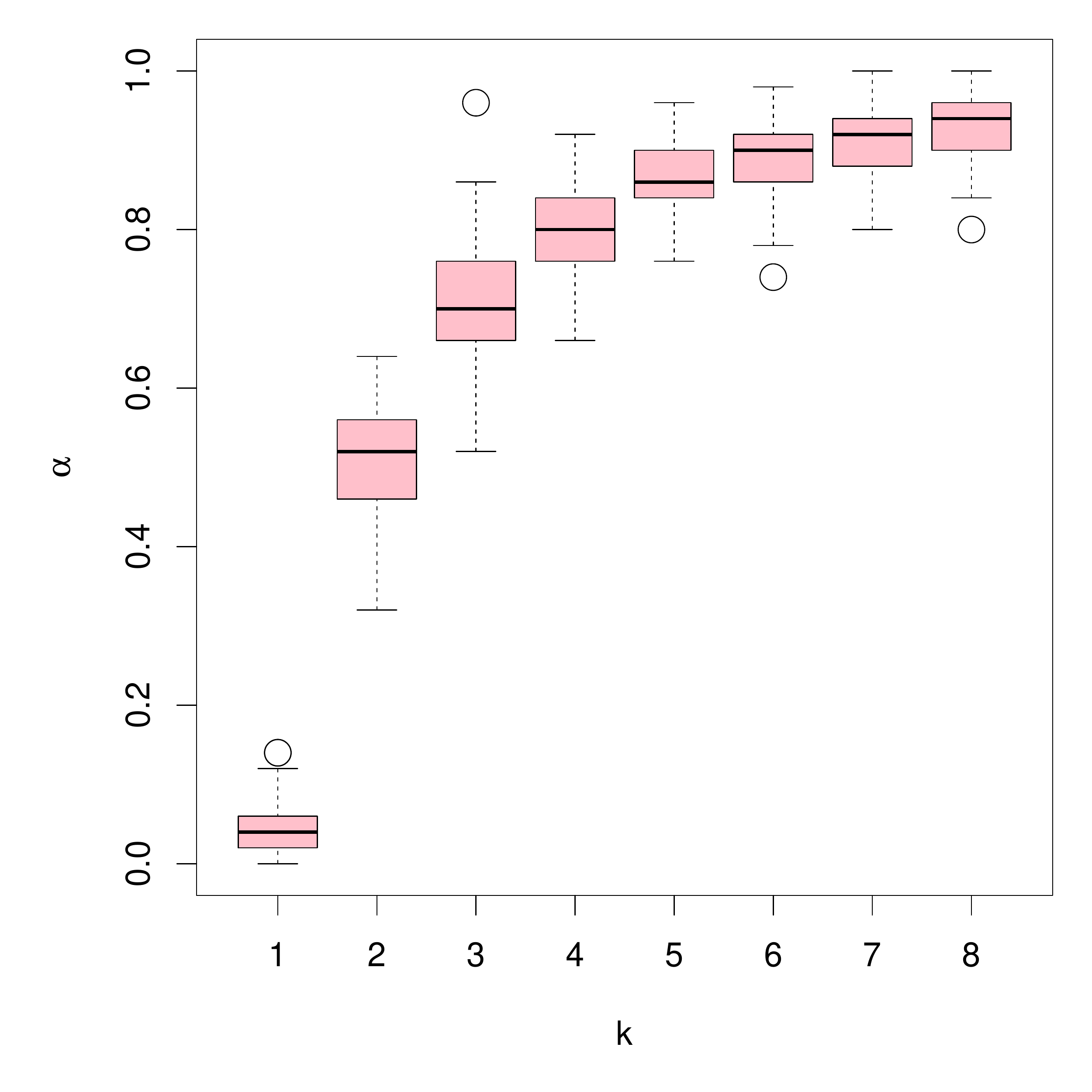}
    \caption{Left panel: marginal correlation which ignore cluster labels (grey) may contradict  inter-cluster correlation which uses clustering labels (blue and red). Right panel: Empirical type I error probability $\alpha$ for testing inter-cluster correlation with zero while ignoring cluster labels deviates from the theoretical value $0.05$  as the number of cluster components increases. Data are simulated  from a mixture of $1 \leq k \leq 8$ standard Gaussian while cluster centres are aligned along $y=x$.}
    \label{Fig:intercluster}
\end{figure}

The test statistic for causal direction relies on a measure of dependence between the error and the predictor. Measuring dependence in heterogeneous populations requires careful considerations as the following simple example shows in Figure~\ref{Fig:intercluster}. Marginal correlation as a measure of dependence which ignores the cluster labels, may contradict the inter-cluster correlation which uses the cluster labels if data are generated from a heterogeneous population. The marginal correlation becomes significantly positive if cluster centres are aligned about the line $y=x$,  and is significantly negative if centres are aligned about $y=-x$, see Figure~\ref{Fig:intercluster} left panel. Consequently, building a test statistic that ignores cluster labels may lead to misleading judgments and affects the test performance, and type I error probability, see Figure~\ref{Fig:intercluster} right panel.

What depicted in Figure~\ref{Fig:intercluster} is essentially Simpson's paradox \citep{simpson1951interpretation} presented  in terms of correlation rather than conditional probability \citep{rucker2008simpson}. The intimate tie between Simpson's paradox and causal inference has been well documented by \citep{pearl2009causality}. This is the main motivation to correct the test of causal direction for heterogeneous data.

\begin{figure}
\centering
		\includegraphics[width=0.7\linewidth]{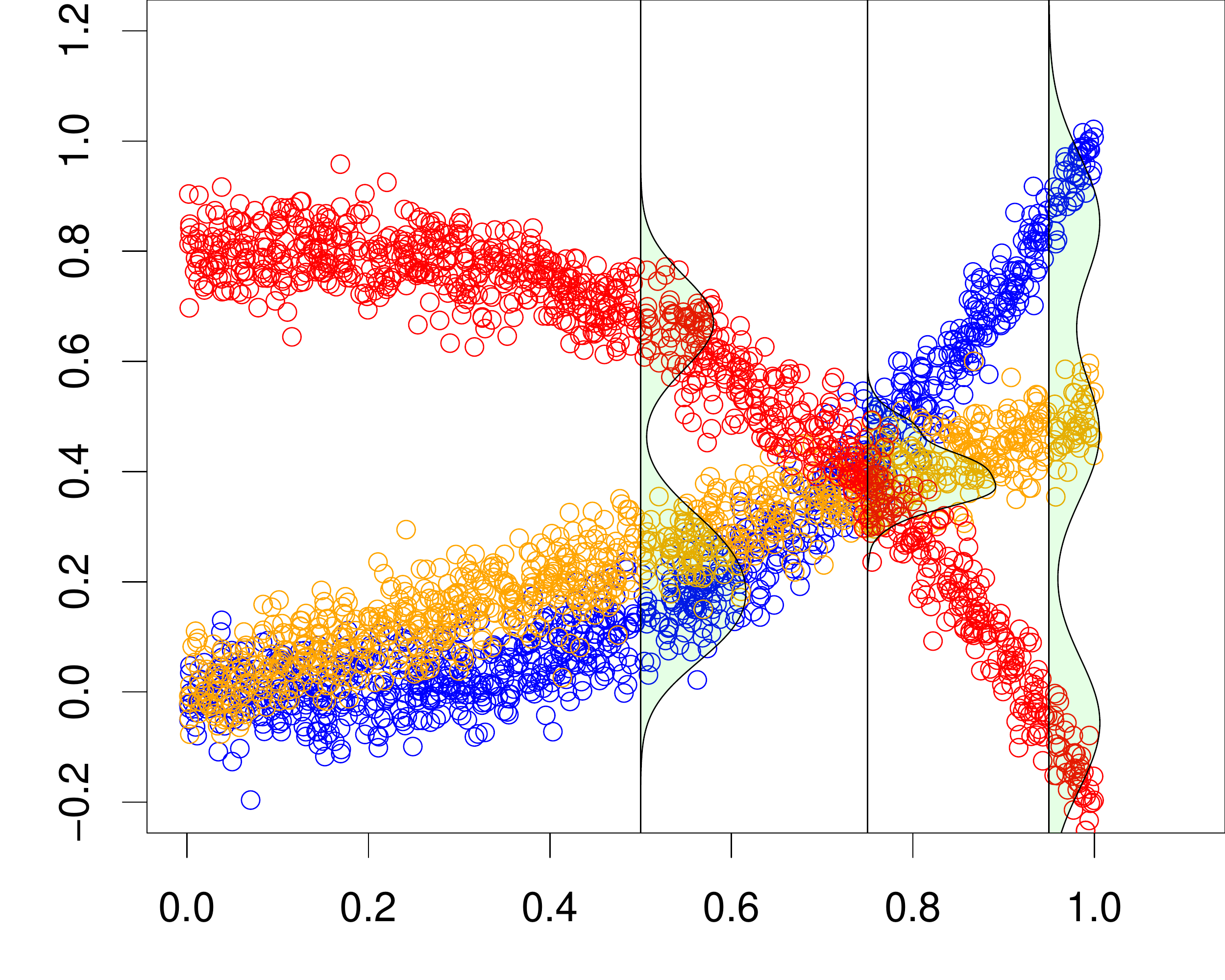}
    \caption{Illustration of from three clusters, $y=x^{3}+\eps$ (blue), $y=0.5x+\eps$ (orange), $y=0.8-x^{3}+\eps$ (red). The causal model imposes a different number of observed components depending on observed $x$, calling for a clustering with flexible component size. Test of causal direction suffers from the same problem in heterogeneous data.}
    \label{Fig:toy}
\end{figure}

Correcting the test statistic requires estimating the cluster labels as well as the number of cluster components. Estimating the number of homogeneous sub-populations is a challenging task. In additive  noise models the observed number of cluster components is heavily affected by the range of the observed input $x$.  Figure~\ref{Fig:toy} provides a visual intuition, where the identified number of cluster components heavily depends on the observed range of $x$. If $x$ is observed about 0.7, a single component emerges. However, observing $x$ about 0.5 or 1.0 changes the number of observed components to 2 or 3 respectively. To this end, we propose a clustering algorithm that labels and the number of components are estimated from the collected data. The estimated labels are used to adjust the causal direction test statistic.


Recently, \citep{liu2016causal} and \citep{Hu_CausalNips_2018} proposed inferring the causal direction on ANMs for discrete and continuous variables respectively. Here we focus on continuous variables. \citep{Hu_CausalNips_2018}  proposed using the $k$-means algorithm on causal parameters and used a predetermined number of sub-populations to overcome data heterogeneity. To the best our knowledge, there is nothing in the literature on correcting the causal direction test statistic after clustering. We speculate that this gap is due to a possible misconception, that is clustering has little effect on causal direction test. We show that this is simply wrong, and the cluster labels play a crucial role in causal direction inference. 

It is evident that the clustering phase is unjustifiable if it does not help inferring the causal direction. Our work builds on  \citep{Hu_CausalNips_2018} and extends it in two directions: i) provides a clustering method with imprecise number of cluster components. ii) uses the clustering information to adjust the test statistic and re-examine the causal direction using clustering labels.

We use a model in which the causal direction of the mixture of ANMs is identifiable, and adopt Partially Observable Gaussian Processes Model for estimation \citep{lawrence2005probabilistic}  proposed in   \citep{Hu_CausalNips_2018}. 

\section{Partially Observed Additive Noise Model}
We assume if $x \to y$, the distribution of $x$ and the function $f$ mapping $x$ to $y$ are independent \citep{janzing2010causal}. We note that $f$ is, for instance, the conditional expectation of $Y$ given $X=x$. The \emph{independence} can perhaps be best understood in a parametric setting where the joint distribution of $(X, Y)$ is known up to finitely many unknown parameters. The independence then means that the marginal distribution of $X$ and the conditional distribution of $Y$ given $X$ do not have any common unknown parameter. In other words, as far as describing the relationship between $X$ and $Y$ are concerned, the conditional distribution of $Y$ given $X$ and the joint distribution of $(X, Y)$ have the same information. So much so that the marginal distribution of $X$ does not have any pertinent information about the conditional distribution of $Y$ given $X$. 

We interpret the independence between the cause and mechanism only through the cluster-specific model parameter $\theta_c$ that captures all properties of the mapping $f$, while $\theta_c$  is independent of the cause $x$. We assume the model is identifiable, i.e. if $x\to y$,  there is no backward additive noise model  $x=g(y;\nu_c)+\eps$ that satisfies  $y \perp\!\!\!\perp (\eps,\nu_c)$.  In other words, if $x$ is  independent of $\theta_c$ in the causal direction, it is  likely that $y$ and $\nu_c$ are dependent in the anti-causal direction \citep[Theorem 1]{Hu_CausalNips_2018}. 

We start the estimation process by projecting a set of centred $n$ dimensional data  $\x=[x_1,\ldots, x_n]^\top$ as the observed cause, and $\y = [y_1,\ldots, y_n]\t,$ as the observed effect onto $d$ hidden dimensions. The projection problem is formalized as the maximization of the Gaussian log-likelihood  
$$\Ell(\K)  = - {dn\over 2}\log (2\pi) - {d\over 2} \log |\K| - {1\over 2} \tr\left(\K^{-1} \y\y\t\right), $$
where $\K$ is the covariance matrix $\K = \pphi\pphi\t +\beta^{-1}\I$, $\beta$ is a positive scale, $\I$ is the identity matrix. The canonical nonlinear feature map $\pphi = [\phi(x_1),\ldots, \phi(x_n) ]\t$  is computed using the kernel trick. 

The latent variable $\theta_i$ is brought in the additive noise model through a concatenated latent predictor $\tilde \x_i\t = [x_i, \theta_i]$ and the Hilbert space is re-defined based on the new vector $\tilde \x_i$.  Therefore, the latent parameters are estimated by maximizing the  Gaussian log-likelihood 
\begin{equation}
\Ell(\ttheta) = -{(d+1)n \over 2} \log(2\pi) - {d+1\over 2} \log |\tilde\K| - {1\over 2} \tr\left(\tilde{\K}^{-1}\y\y\t\right),    
\end{equation} 
where $\ttheta$ is the vector composed of $\theta_i$'s, $\tilde \K = \tilde\PPhi \tilde\PPhi\t$, and  $\tilde\PPhi_{n\times (d+1)} = [\phi(\tilde \x_1), \ldots, \phi(\tilde \x_N)].$ The parameter vector $\ttheta$ appears in $\tilde \K$ through $\tilde \x$. In our developments we focus on univariate $\theta_i$, but the methodology is general and is valid for multivariate projection as well.
This approach re-formalizes the additive model $y= f(x,\theta) + \eps$  in terms of the augmented variable $ y =f(\tilde \x) + \eps.$ However, still the log-likelihood of an ill-defined model, because  $x$  and $\theta$ should be independent. 

The Hilbert-Schmidt independence criterion (HSIC) measures the dependence between observations of a pair of random variables by   projecting them onto  the reproducing kernel Hilbert space. The empirical HSIC is 
\begin{equation}
\mathrm{HSIC} = {1\over n^2} \tr(\K\H\L\H),
\end{equation}
where $\K$ is the kernel element of $x, k(x_i,x_{i'}),$ $\L$ is the kernel element of $y, l(y_i, y_{i'})$, $\H=\I- {1\over n } \one\one\t$ and $\one$ is the unit vector of size $n$.

The independence between $x$ and $\theta$ is encouraged by adding HSIC as a regularizer to the log-likelihood term through the regularization constant $\lambda>0$ 
\begin{equation}
\Jll (\ttheta) = \Ell (\ttheta) - \lambda \log \HSIC(\ttheta).    
\label{eq:Jll}
\end{equation}
Causal parameters are estimated by $\hat{\ttheta} = \argmax \Jll(\ttheta)$ using scaled conjugate gradient maximization \citep{Hu_CausalNips_2018}.

We first estimate the model parameters for each subject $\theta_i, i=1,\ldots, n$, we then cluster $\theta_i$ while the number of clusters $k$ is imprecise. This maps $\theta_i$ to $\theta_c, c=1,\ldots,k$ while $k$ varies $K-\Delta\leq k \leq K+\Delta$ for a given $\Delta$. The cluster component range $\Delta$ has little effect on the performance of the clustering algorithm and is determined by the computational budget.

\section{Clustering Method}\label{section:clust}

The proposed clustering method works via combining additive noise models \citep{hoyer2009nonlinear} with product partition models \citep{hartigan1990partition}. We provide more details about the  intuition behind the clustering method below. This intuition provide some insights about how to estimate the number of cluster components.

\begin{figure}
  \begin{center}
    \begin{tikzpicture}
    \node[latent,fill=yellow] (theta) {$\theta\mid \mu$};
    \node[latent, right=of theta, fill=cyan] (f) {$f(x\mid\ttheta,\z)$};
    \node[latent, above=of f,fill=yellow] (x) {$x$};
    \node[latent, right=of f,fill=orange] (y) {$y$};
    \node[latent, above=of x,fill=green] (px) {$p(x)$};
    \node[latent, above=of theta,fill=yellow] (pt) {$\mu$};
    \node[latent, above=of y,fill=yellow] (e) {$\eps$};
    \node[latent, above=of e, fill=green] (pe) {$p(\eps)$};
    \node[latent, above=of pt,fill=green] (pm) {$p(\mu)$};
    \node[latent, below=of f,fill=yellow] (z) {$\z$};

\node at (3,1) [draw,rectangle,minimum width=4cm,minimum height=7.5cm,color=blue] {};
\node at (3,5) []  (t1) {\color{blue}\small  $f(x,y\mid\ttheta, \z)$};

    \edge {px} {x}; %
    \edge {pt} {theta};
    \edge {x,theta,z} {f};
    \edge {pe} {e};
    \edge {e,f} {y};
    \edge {pm} {pt};
    \end{tikzpicture}
  \end{center}
  \caption{The generative process of the proposed causal model $f(x\mid\ttheta)$ and how it interacts with the clustering model $f(x,y\mid\ttheta, \z)$ of equation \eqref{eq:prodpart}.}
  \label{fig:dag}
  \vspace{-2mm}
\end{figure}
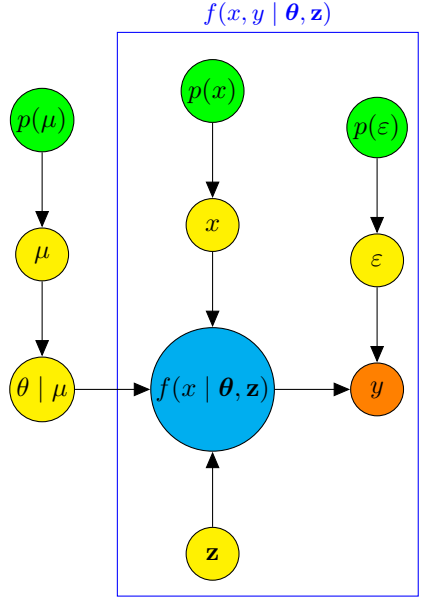

The clustered version of the additive noise models is composed of several additive noise models of the same causal direction \citep{Hu_CausalNips_2018}. Causal models are developed given data labels which are to be estimated, either mutually or after estimation of causal model parameters. Following  \citep{Hu_CausalNips_2018} we propose the latter approach since it is computationally less demanding.

A cluster additive noise model is a set of causal models of the same causal direction between two continuous random variables  $x$ and $y$ with realizations $x_i$ and $y_i$, $i=1,\ldots, n$
    \begin{align}
    	y_i = f(x_i\mid\theta_c) + \eps_i,
        \label{mix_anm}
    \end{align}
where $x$ denotes the cause, $y$ denotes the effect, $f$ is a smooth  function, nonlinearity parameters $\theta_c$ that parametrize the smooth function $f$, and $\eps$ is the statistical noise.

In model \eqref{mix_anm} we assume  
\begin{enumerate}
    \item $x_i\sim p(x), \theta_c \sim p(\theta)$ are independently and identically drawn from a Gaussian distribution. 
    \item The statistical error is independent of the covariates and clustering parameters $\eps_i\perp\!\!\!\perp x_{i'}$ and $\eps_i \perp\!\!\!\perp \theta_c$. 
    \item The clustering parameters $\theta_c$  independently and identically drawn from a Gaussian distribution.
\end{enumerate} 

The difference between traditional causal models and cluster causal models is the way that causal parameters $\theta$ parametrize the smooth function $f$. The nonlinearity parameter $\theta_c$ is drawn randomly from a probabilistic  model independently. In other words, a set of independent generating mechanisms is assumed for each sub-population through $\theta_c$. Our formulation is slightly different from  \citep{Hu_CausalNips_2018} that assumes the causal parameters $\theta_c$ are drawn from a fixed set,but we assume they are generated from an independent Gaussian distribution. Although this modification seems minor, it plays a major role in attaching the causal model to the clustering algorithm in Figure~\ref{fig:dag} and allows to aggregate the causal test statistic through the cluster independence assumption. This model is inspired by commonly encountered situations where the data generating process from one independent trial to another may be different due to the influence of uncontrollable factors.

We first re-formalize probabilistic clustering model through Bayesian regression and derive the clustering algorithm using this model. This viewpoint allows us use the marginal  posterior as an estimation tool for the cluster component size.

Assume the following Bayesian regression for the latent parameters
\begin{eqnarray}
 \ttheta \mid \mmu &\sim&  \N(\Z\mmu, \sigma^2 \I) \nonumber \\
 \mmu &\sim& \N(\bar{\ttheta} , \kappa(\tau\sigma)^2\I)\label{eq:gaussmodel},
\end{eqnarray}
where $\sigma^2$ is the common within-cluster variance,  $\tau^2$ is the between-cluster to within-cluster variance ratio, and $\bar\ttheta=[\bar\theta_1,\ldots,\bar\theta_k]$ is the vector of the cluster averages.  The over-dispersion parameter  $\kappa> 1$  controls the prior information, i.e.  a large $\kappa$ value gives a flat prior with minimal prior information about the parameters. 
This model is a sort of \emph{empirical Bayes} in which the data statistic is utilized to parametrize the prior.

 We adopt the \emph{product partition} model \citep{hartigan1990partition} for clustering, i.e.
    \begin{equation}
        f(\x, \y \mid \ttheta, \z) = \prod_{c=1}^k \prod_{\{i\mid z_i=c\}} f(x_i, y_i\mid \theta_i)
        \label{eq:prodpart}
    \end{equation}
    The clustering mechanism adds an unobserved  label   $z_i$ to each observation, i.e.    
 $\left(x_i, y_i, z_i\right)$  or equivalently $(\theta_i,z_i)$ in which $z_i \in \{1,2,\dots,k\}$ is the label and  $k$ is the uncertain number of sub-populations. 
    The clustering method relies on $\theta_i \in \real$ which is the key to distinguish between different generating mechanisms. Note that for an identifiable mapping $f_\theta$, one can directly cluster generating mechanisms by clustering $\theta_i$'s. Therefore, $\theta_c$ in  \eqref{eq:prodpart} is equivalent to the pair $(\theta_i, z_i=c)$.

    A practical causal cluster model should uncover  the unknown number of cluster components $k$ as well as the unobserved  label $z_i$. We therefore focus on devising an algorithm that allows for clustering $\theta_i$ with a flexible component size  $k = \max(z_i) \in \{K-\Delta,\ldots, K+\Delta\} $, given positive integers $K$ and $\Delta$, $K>\Delta+1$. The algorithm looks like a simple extension of $k$-means, but the inspiration comes from a probabilistic clustering that satisfies certain conditions to guarantee convergence.

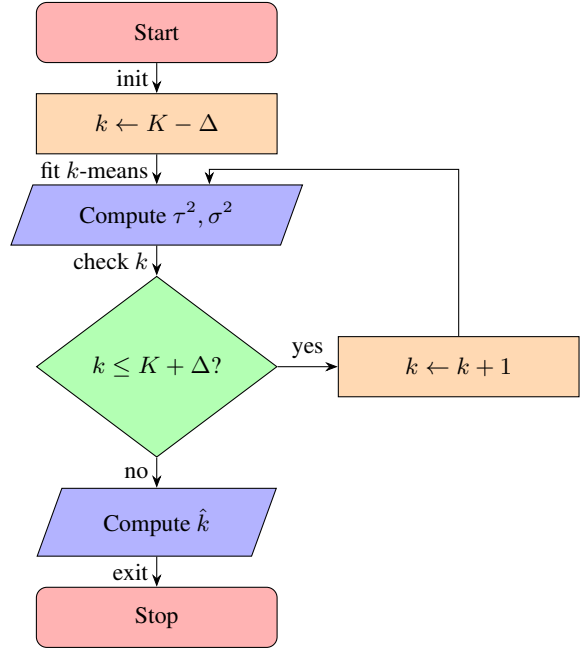
\begin{figure}\centering
 \small \begin{tikzpicture}[
    node distance = 4mm and 8mm,
      start chain = A going below,
      base/.style = {draw, minimum width=32mm, minimum height=8mm,
                     align=center, on chain=A},
 startstop/.style = {base, rectangle, rounded corners, fill=red!30},
   process/.style = {base, rectangle, fill=orange!30},
        io/.style = {base, trapezium, 
                     trapezium left angle=70, trapezium right angle=110,
                     fill=blue!30},
  decision/.style = {base, diamond, fill=green!30},
  every edge quotes/.style = {auto=right}]
                    ]
\node [startstop]       {Start};            
\node [process]         {$ k \leftarrow K-\Delta$};
\node [io]              {Compute $\tau^2,\sigma^2$};
\node [decision]        {$k\leq K+\Delta$?};
\node [io]         {Compute $\hat k$};
\node [startstop]         {Stop};             
\node [process,                             
       right=of A-4]    {$k \leftarrow k+1$};
\draw [arrows=-Stealth] 
    (A-1) edge["init"]          (A-2)
    (A-2) edge["fit $k$-means"]    (A-3)
    (A-3) edge["check $k$"]       (A-4)
    (A-4) edge["no"]            (A-5)
    (A-5) edge["exit"]          (A-6)
    (A-4) edge["yes"']          (A-7)       
    (A-7) |- ($(A-2.south east)!0.5!(A-3.north east)$)
          -| ([xshift=7mm] A-3.north)
    ;
  \end{tikzpicture}
  \caption{The condenced flow chart of the proposed clustering algorithm while $k$-means is used as a  subroutine.}
    \label{fig:flowchart}
\end{figure}

\begin{enumerate}
\item Initialization: Set $K, \Delta$, initialize $\z$.
    \item Run $2\Delta+1$ clustering chains in parallel $k\in\{K-\Delta,\ldots, K+\Delta\}$
    \item For each chain of size $k$
    \begin{enumerate}
        \item[3.1)] centre update:  $\mu_c = \bar\theta_c$

        \item[3.2)] label update: $z_i = \argmin_c  {\vert\theta_{i} - \mu_{c}\vert} .$

    \end{enumerate}
        \item[4.]  Within-cluster variance computation: $$\sigma^2 = {1\over N }\sum\limits_{c=1}^k \sum\limits_{\{i\mid z_i=c\}}^{n_c} ( \theta_{i} -\bar\theta_c )^2.$$
        \item[5.] Between-to-within variance ratio computation: $$\tau^2 = {1\over k\sigma^2} \sum_{c=1}^k ( \bar\theta_c -{\bar \theta} )^2$$
\item[6.] Component size estimation: $\hat k = \argmax\limits_{k} \ell(\z\
    \mid \ttheta,k).$
\end{enumerate}
We iterate between 3.1 and 3.2 until convergence and ultimately at step 5 report the labels  $\z$ with $k$ that maximize  

\begin{equation}
\ell(\z\mid\ttheta,k) = -{n\over 2 } \log 2\pi \sigma^2 
 -{1\over 2 \sigma^2} \sum_{c=1}^k\sum_{\{i\mid z_i=c\}} ( \theta_{i} - \bar \theta_{c})^2 
 - {1 \over 2} \sum_{c=1}^k \log(\kappa\tau^2 n_c +1),  
\label{eq:clustloglike}
\end{equation}
see Figure~\ref{fig:flowchart} that visualizes the algorithm in a condensed flowchart.

The clustering algorithm resembles the $k$-means to a great extent and only adds a few more steps to estimate the cluster component size using the marginal log likelihood $\ell.$ The clustering hyperparameter $\kappa$ is a sort of over-dispersion of cluster centres. Our experiments show $ \kappa=e^{kn/4}$ is a good choice. The computational complexity of the clustering algorithm is $\mathcal O(\Delta NK).$  The following result (Theorem ~\ref{theo:conv}) shows that a stochastic version of the proposed clustering algorithm converges to a stationary distribution.  


\begin{thrm}
\label{theo:conv}
Suppose $\sigma^2$ and $\tau^2$ are given. 
\begin{enumerate}
\item
cluster centre update: sample from a Gaussian distribution with mean $$\mu_c \sim \N\left( \bar\theta_c (1 + {1\over n_c\kappa\tau^2})^{-1}, \sigma^2(n_c+{1\over \kappa\tau^2})^{-1}\right).$$
\item
cluster label update: sample from Multinomial distribution with probability mass 
  $$z_i\sim \Pr(z_i=c)=  {\phi\left( {\theta_i - \mu_c \over \sigma  }\right)\over  \sum\limits_{c=1}^{k} \phi\left( {\theta_i - \mu_c \over \sigma  }  \right)}, c=1,\ldots, k$$ in which $\phi(\cdot)$ is a standard Gaussian density.
  \item
  cluster component update: sample  from Multinomial distribution with probabilities proportional to $ \exp\{\ell(\z\
    \mid \ttheta,k)\}$ of \eqref{eq:clustloglike}. 
\end{enumerate}
Then the proposed clustering algorithm converges to the stationary distribution  
$$p(\z\mid\ttheta)\propto \sum_{k=K-\Delta}^{K+\Delta} \exp\{\ell(\z\mid \ttheta,k)\}$$\\
\end{thrm}
See Appendix for the proof.

This probabilistic clustering method resembles $k$-means and cluster size estimation  resembles BIC scoring \citep{schwarz1978estimating}. 
There has been various attempts to use BIC for cluster component selection.  Several authors including \citep{pelleg2000x} studied $k$-means with BIC scoring and report that BIC over-estimates the number of components. Our scoring is developed for the clustering context, in which $k$-means matches the setting of Bayesian linear regression: labelling update is equivalent to design matrix estimation, and mean update is equivalent to the coefficient estimation. Our formulation shows that the conventional BIC scoring is inappropriate for $k$-means and requires proper generalization. It also  shows the marginal posterior scoring with $\ell$ becomes BIC scoring of \citep{pelleg2000x} if clusters are balanced $n_c = n_{c'}, c\neq c'$, and  $\kappa\tau =1$.



\section{Test Statistic Adjustment}
In additive noise causal models,  $\HSIC$ is proposed to draw conclusions about the causal direction \citep{Hu_CausalNips_2018}. This test statistic is, however, designed for situations in which samples are coming from a homogeneous population, and is highly sensitive to departures from this assumption. Theorem ~\ref{theo:HSIC} presents the asymptotic distribution of an adjusted empirical HSIC to infer the causal direction in a heterogeneous case for a given set of clustered data.
\begin{thrm}
\label{theo:HSIC}
Define $\HSIC_c = {1\over n^2_c} \tr(\K_c \H_c \L_c \H_c) $ to be the cluster-specific empirical $\HSIC$ statistic.  The aggregated  test statistic $ t= \sum_{c=1}^k n_c \HSIC_c$ converges in distribution to $\sum_{l=1}^{\infty} \lambda_l z^2_l$, where $z_l$, $l=1, 2, \cdots$ are independent standard Gaussian random variables and $\lambda_l$ $l=1, 2, \cdots$ are non-negative  constants. 
\end{thrm}

The result of Theorem ~\ref{theo:HSIC} reduces to the homogeneous case of \citep{gretton2005kernel} if $k=1$.  The  theoretical quantile of the test statistic $t$ can be  calculated using the Gamma basis   $\alpha = {\mu^2\over \sigma^2}$ and $ \beta = {\sigma^2\over \mu}$ with $  \mu =  \sum_{c=1}^k n_c \mu_c, \sigma^2 = \sum_{c=1}^k n_c^2 \sigma^2_c$  \citep{wood1993saddlepoint},
in which $\mu_c$ and $\sigma^2_c$ are the cluster-specific theoretical $
\HSIC$ mean and variance. 

\section{Application}
The T\"uebingen cause-effect pairs \citep{mooij2016distinguishing} is a well-known benchmark in the context of causal direction detection \footnote{\url{https://webdav.tuebingen.mpg.de/cause-effect/}}.   The database includes 41 data sets arranged in 108 pairs $(x,y)$ with a known causal direction identified for each pair, either $x\to y$ or $y\to x$.

\subsection{Life Expectancy Data}
\begin{figure}
\includegraphics[width=.55\textwidth]{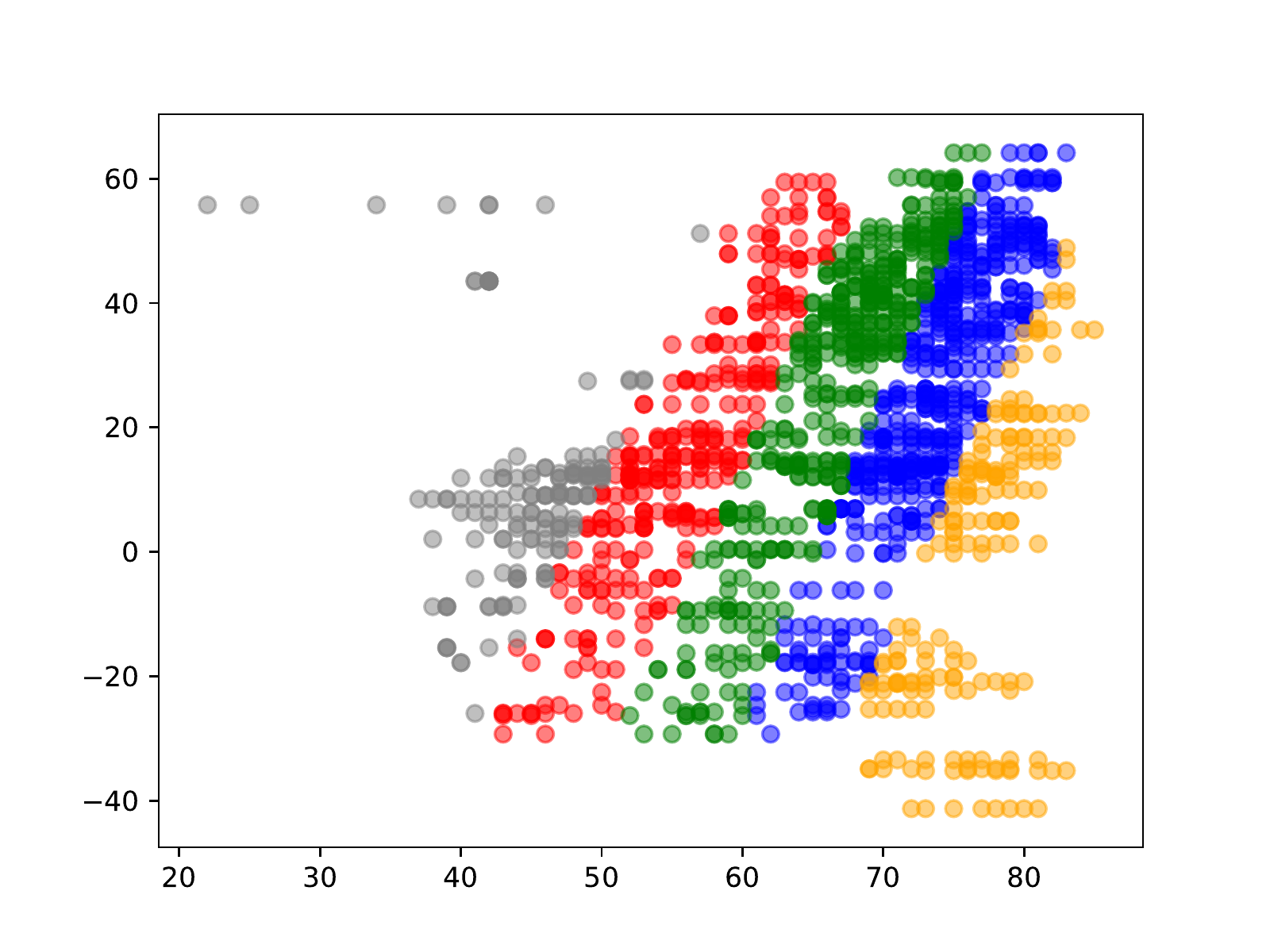}
\includegraphics[width=.55\textwidth]{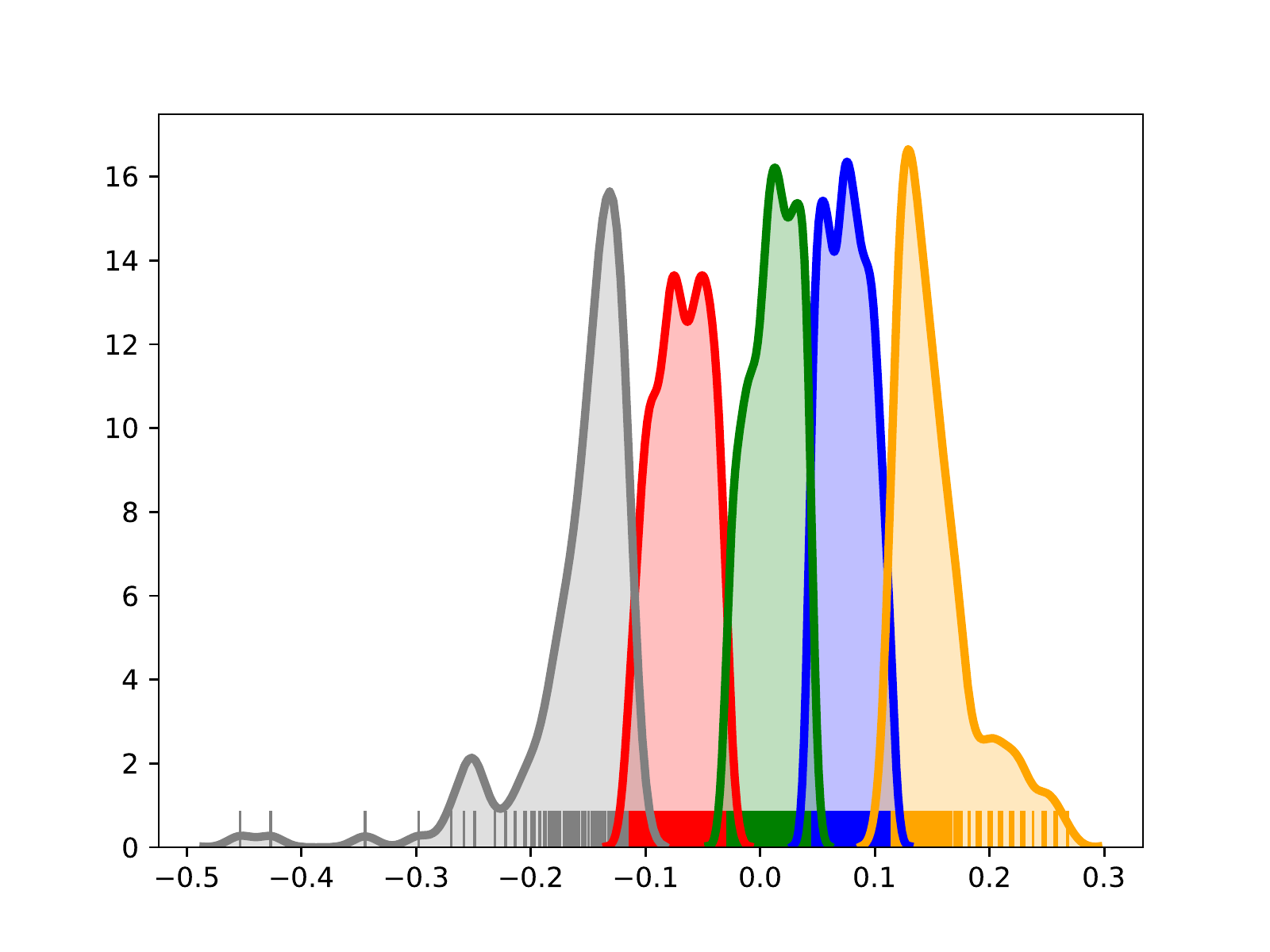}
\caption{Top panel: the UN life expectancy dataset is clustered to $k=5$ using our clustering algorithm. Bottom panel: the density plot of the causal parameters $\theta_i$.}
\label{fig:example}
\end{figure}

First we explore the effect of the number of clusters on the test statistic for the UN life expectancy data by concatenating pairs 56--63 of T\"uebingen cause-effect pairs. 
Figure~\ref{fig:example} (left panel) shows the scatter plot of UN data $x$: life expectancy versus $y$: latitude; note that the true causal direction is $y\to x$. 

Data are generated from multiple sources, so we expect that data homogeneity assumption fail to hold. The scatter plot in Figure~\ref{fig:example} confirms this visually.  

We compute the causal parameters by maximizing the log likelihood \eqref{eq:Jll} with $\lambda=50$ given the true causal direction $y\to x$.  The test statistic without adjustment is $14.90$  and its theoretical $5\%$ quantile is $0.60$, so it mistakenly rejects the null. The statistic after adjustment using clustering labels with $k=2,3,4$ still rejects the true direction but with a larger $p$-value. This is aligned with our observation in the simulated mixture example in Figure~\ref{Fig:intercluster} (left panel), i.e. for a large number of components the type I error is more affected, and we expect to see the effect of test statistic correction specially for large $k$.

In this example $k=5$ gives the adjusted test statistic $t=1.73$ with the $5\%$ theoretical quantile $2.21$, so infers the causal direction correctly. The  direction is inferred correctly also for $k\geq 6$.  Our clustering algorithm  estimated $7$ clusters.

\subsection{Multiple Cause-Effect Pairs}

Next we  check the performance of our method on all data pairs as well. Following \citep{Hu_CausalNips_2018} we exclude pairs 12,17, 47, 52, 53, 54, 55, 70, 71, 101, and 105. Additionally we excluded pairs 73, 106, and 68 that include  outliers. These outliers yielded singleton clusters troubled the computation of the cluster specific test statistic numerically. 

\begin{figure}
\centering
\includegraphics[width=.5\textwidth]{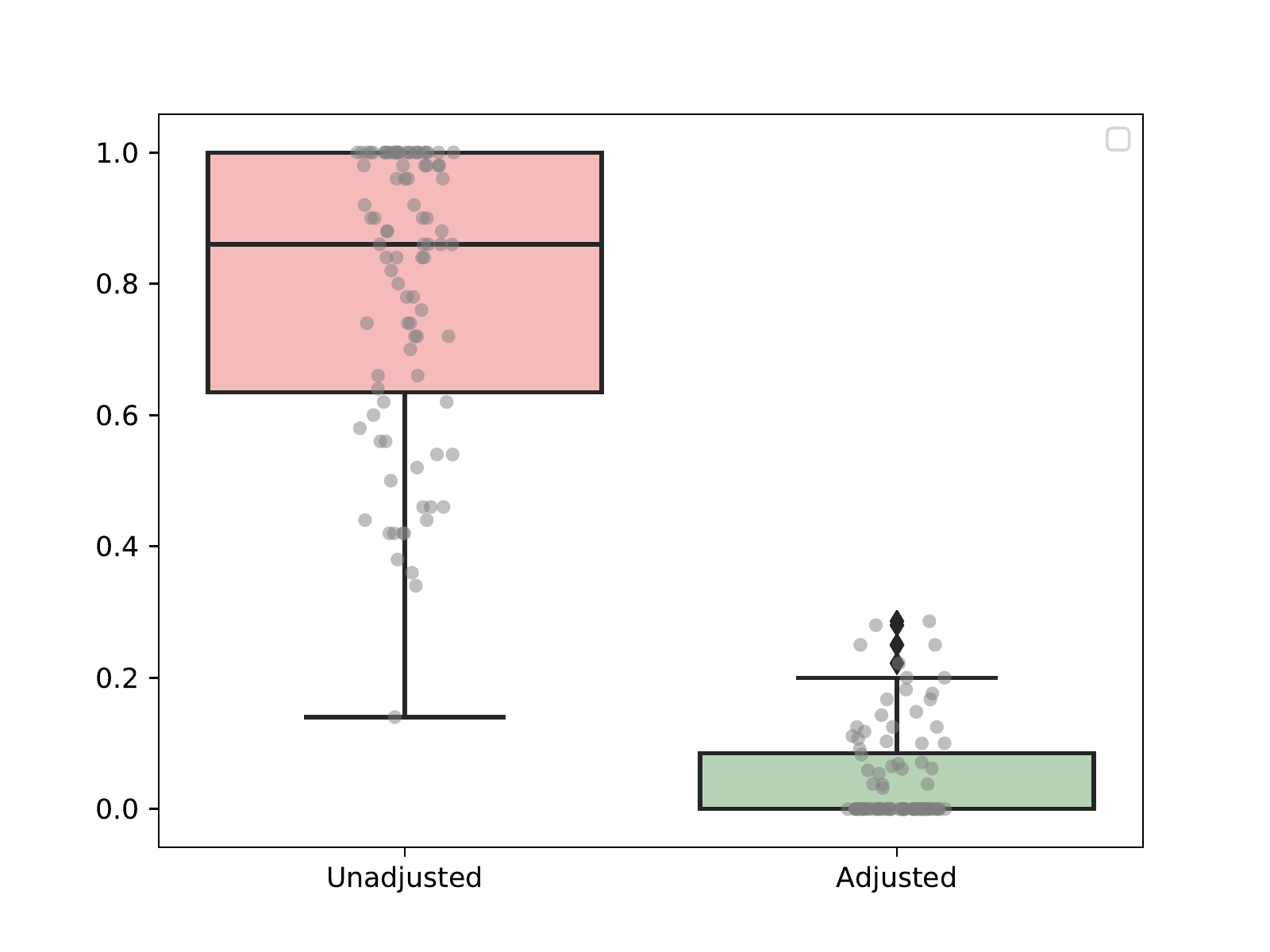}
\caption{Boxplot of type I error probability  of T\"uebingen cause-effect pairs unadjusted (left) versus adjusted (right). The nominal type I error remains on average around the nominal value $0.05$ after adjustment.}

\label{fig:boxplot}
\end{figure}

We sample 90 data from each pair and repeat this process 50 times  independently. Then we estimate the causal parameters by maximizing the log likelihood \eqref{eq:Jll} with $\lambda=50$ for $x\to y$ and $y\to x$ directions. We choose $k = K\pm 2$ with a visually appealing $2 \leq K \leq 6$ for each data set.  We used the clustering labels to adjust the $\HSIC$ statistic while running our clustering algorithm. Figure~\ref{fig:boxplot} shows the boxplot of type I error.  The total error probability (type I error + type II error) remain equal. Theoretically the type I error probability must remain under control at about the significance level. However, the mean of type I error probability for unadjusted statistic is 0.796 while using the adjusted method it is 0.048. The latter preserved the nominal type error probability 0.05.
We see a similar behaviour in the simulated toy  example of Figure~\ref{Fig:intercluster} in which the type I error probability is exceedingly higher than the nominal value for large number of components $k$.

\subsection{Disease Symptom Data}
\begin{figure}
\centering
\includegraphics[width=.3\textwidth]{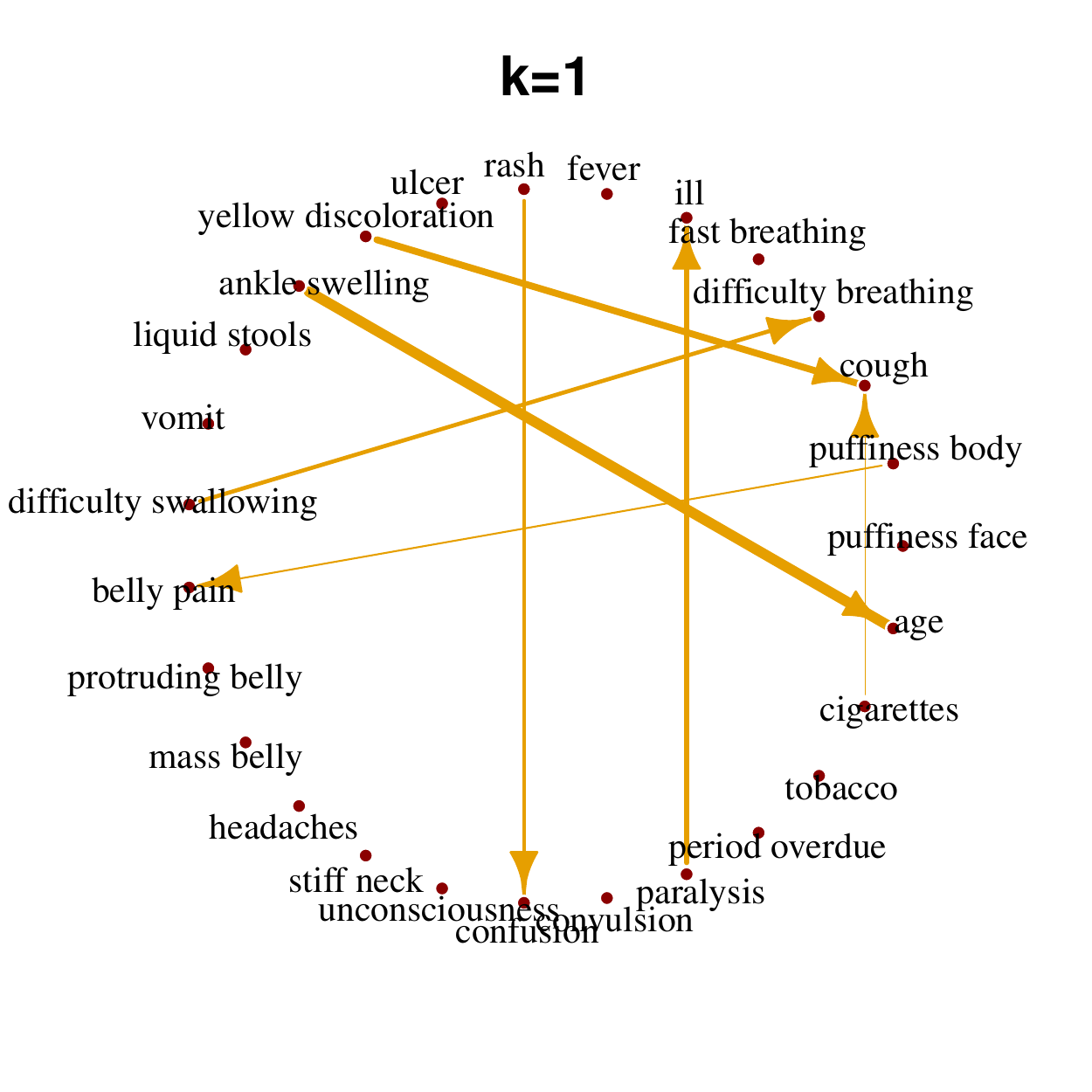}~~
\includegraphics[width=.3\textwidth]{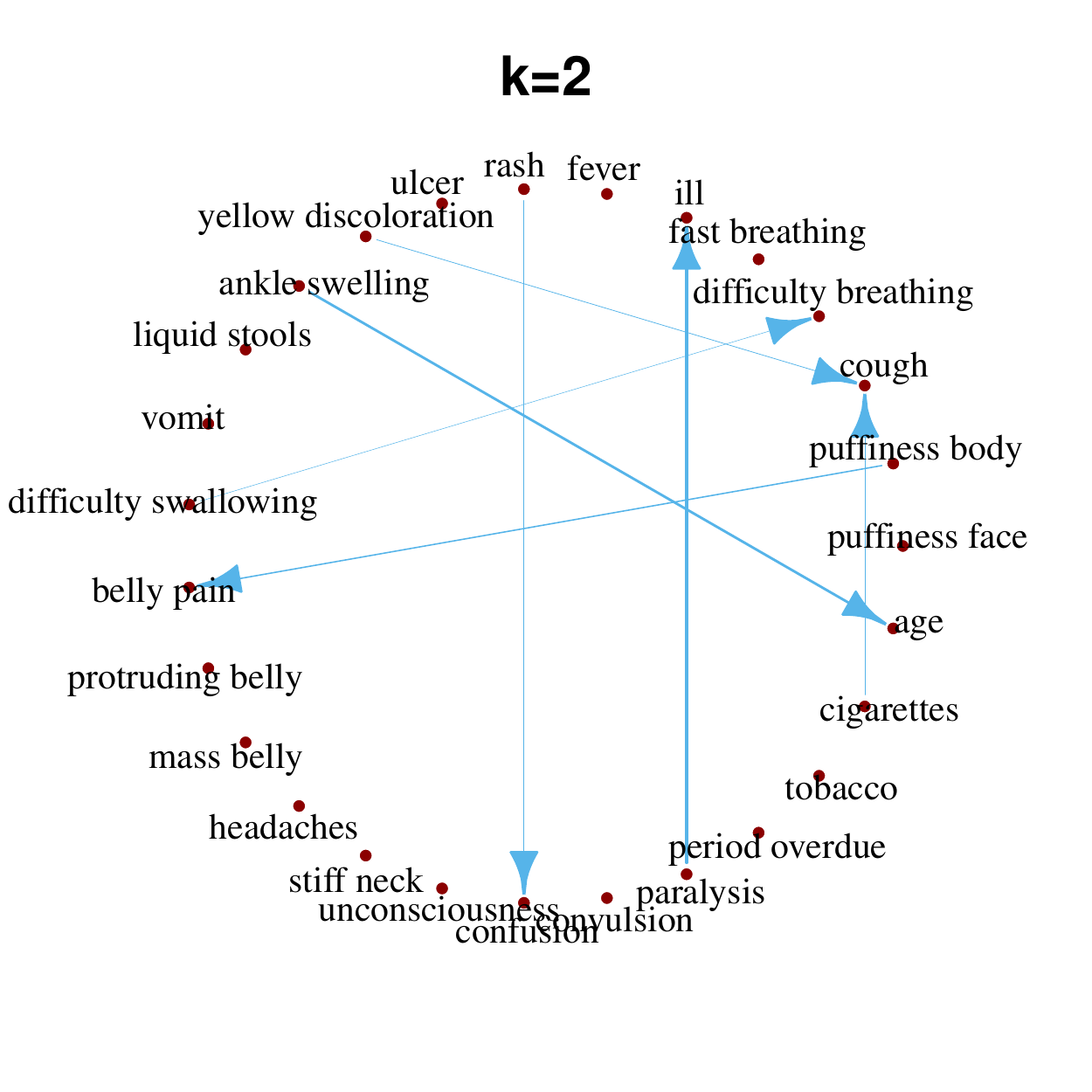}~~
\includegraphics[width=.3\textwidth]{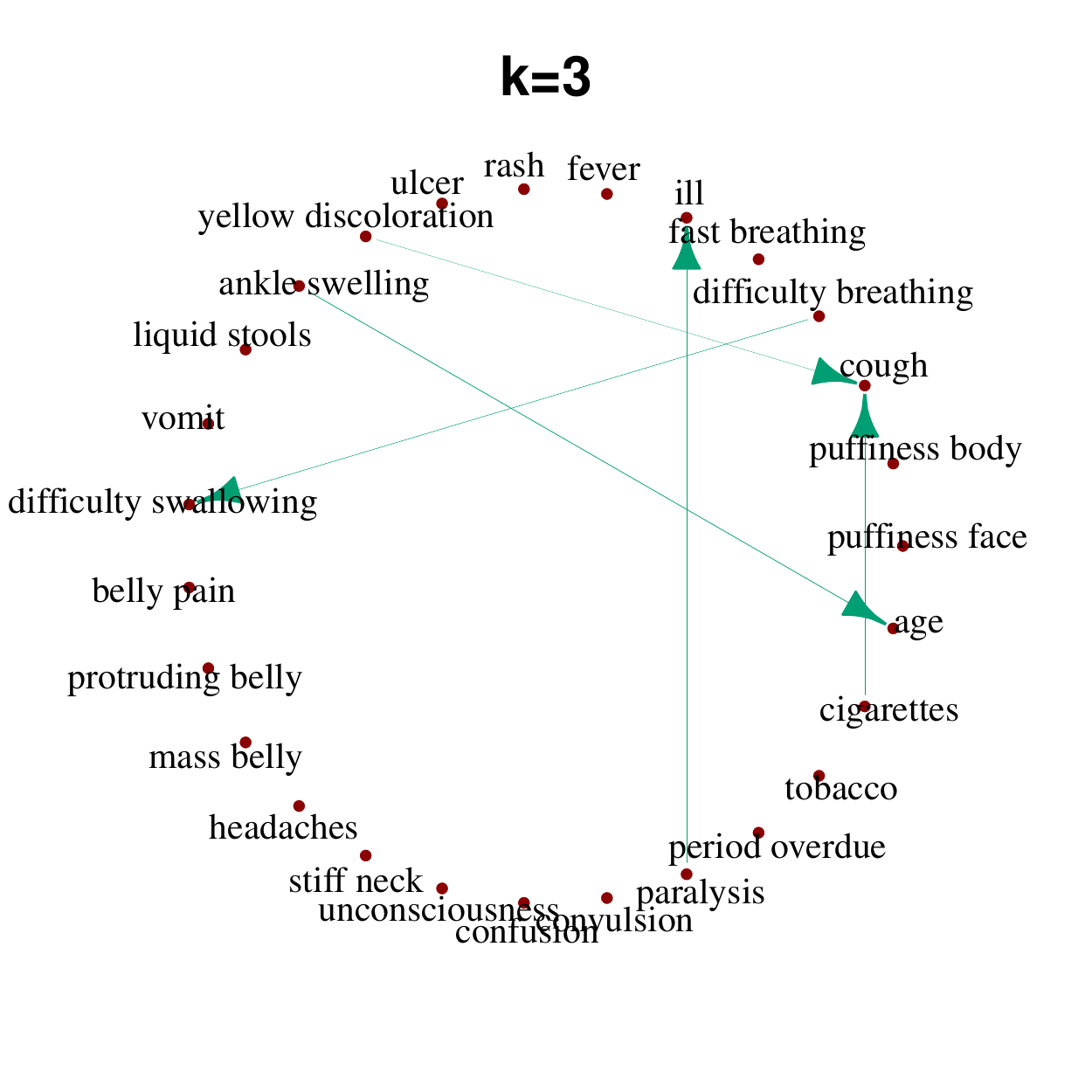}

\caption{Causal direction for different clusters, homogeneous case with a single cluster $k=1$ (left panel), heterogeneous case with two clusters $k=2$ (middle panel), and with three clusters $k=3$ (right panel). Edges tend to weaken by increasing $k$. In $k=3$, some edges disappear, and a directed edge is reversed.}

\label{fig:causaldirection}
\end{figure}

We analyze the stroke subset of verbal autopsy survey \citep{Murray_VerbalAutopsyData_2011} benchmark data available in \texttt{openVA} R package. Stroke measures 27 cause and symptoms ranging from vague symptoms such as \emph{ill} to specific symptoms such as \emph{vomit}.  Inference about the association between variables are made using the spike and slab model of \citep{Li_BGLASSO_2019} which only discovers important associations between variables. In contrast, our method can enhance the analysis further by finding the causal direction between the dependent variables with or without the homogeneity assumption, i.e. $k=1$ or $k > 1$. 
Scatter plot of data pairs clearly indicates that homogeneity is not a viable assumption. Figure~\ref{fig:causaldirection} confirms the inferred direction depends on the number of clusters.

\section{Conclusion}
We showed that heterogeneity can severely affect causal direction inference. In fact, the distribution of the test statistic deviates from the theoretical distribution obtained under the homogeneity assumption. This deviation considerably affects the type I error probability. The $\HSIC$ test statistic used in causal direction identification relies heavily on the homogeneity assumption. It requires proper adjustment using clustering labels when there are reasons to believe that homogeneity assumption is not tenable. To adjust this test statistic, we first developed a clustering method in the context of additive noise models that allows flexible number of clusters. We then used the estimated clustering labels to adjust the test statistic of causal direction for heterogeneous  data.

It is well-known in hypothesis testing that  type I error and type II error work in opposite direction, e.g. decreasing type I error leads to an increase in type II error. For instance, in likelihood ratio tests for composite hypotheses $\alpha + \beta \leq 1$. Total error improves only in a local neighbourhood of the null hypothesis only if the sample size increases.   Our method adjusts type I error probability while keeping the total error in the same order. In some of our experiments it even improves the total error so this adjustment builds an empirically more efficient test. The unadjusted HSIC test has a wrong asymptotic distribution when homogeneity assumption fails to hold. This is why, the type I error deviates from its nominal level. 


To implement our proposed clustering algorithm we assumed that $K$ and $\Delta$ are known. This assumption was only made to facilitate computation. The method, however, works for any value of $K$ and $\Delta$. This restriction can be relaxed by setting $K-\Delta=1, K+\Delta=N$ if the computational power allows. This modification makes the clustering algorithm $\mathcal O(N^3)$.




\bibliography{ref-causal}

\begin{thebibliography}{29}
\expandafter\ifx\csname natexlab\endcsname\relax\def\natexlab#1{#1}\fi
\providecommand{\url}[1]{\texttt{#1}}
\providecommand{\href}[2]{#2}
\providecommand{\path}[1]{#1}
\providecommand{\DOIprefix}{doi:}
\providecommand{\ArXivprefix}{arXiv:}
\providecommand{\URLprefix}{URL: }
\providecommand{\Pubmedprefix}{pmid:}
\providecommand{\doi}[1]{\href{http://dx.doi.org/#1}{\path{#1}}}
\providecommand{\Pubmed}[1]{\href{pmid:#1}{\path{#1}}}
\providecommand{\bibinfo}[2]{#2}
\ifx\xfnm\relax \def\xfnm[#1]{\unskip,\space#1}\fi
\bibitem[{Dehkharghani et~al.(2014)Dehkharghani, Mercan, Javeed \&
  Saygin}]{dehkharghani2014sentimental}
\bibinfo{author}{Dehkharghani, R.}, \bibinfo{author}{Mercan, H.},
  \bibinfo{author}{Javeed, A.}, \& \bibinfo{author}{Saygin, Y.}
  (\bibinfo{year}{2014}).
\newblock \bibinfo{title}{Sentimental causal rule discovery from twitter}.
\newblock {\it \bibinfo{journal}{Expert Systems with Applications}\/},  {\it
  \bibinfo{volume}{41}\/}, \bibinfo{pages}{4950--4958}.
\bibitem[{Fisher(1926)}]{fisher1926arrangement}
\bibinfo{author}{Fisher, R.~A.} (\bibinfo{year}{1926}).
\newblock \bibinfo{title}{The arrangement of field experiments}.
\newblock In {\it \bibinfo{booktitle}{Breakthroughs in statistics}\/} (pp.
  \bibinfo{pages}{82--91}).
\newblock \bibinfo{publisher}{Springer}.
\bibitem[{Greenland et~al.(1999)Greenland, Pearl \&
  Robins}]{greenland1999causal}
\bibinfo{author}{Greenland, S.}, \bibinfo{author}{Pearl, J.}, \&
  \bibinfo{author}{Robins, J.~M.} (\bibinfo{year}{1999}).
\newblock \bibinfo{title}{Causal diagrams for epidemiologic research}.
\newblock {\it \bibinfo{journal}{Epidemiology}\/},  (pp.
  \bibinfo{pages}{37--48}).
\bibitem[{Gretton et~al.(2005{\natexlab{a}})Gretton, Bousquet, Smola \&
  Sch{\"o}lkopf}]{gretton2005measuring}
\bibinfo{author}{Gretton, A.}, \bibinfo{author}{Bousquet, O.},
  \bibinfo{author}{Smola, A.}, \& \bibinfo{author}{Sch{\"o}lkopf, B.}
  (\bibinfo{year}{2005}{\natexlab{a}}).
\newblock \bibinfo{title}{Measuring statistical dependence with hilbert-schmidt
  norms}.
\newblock In {\it \bibinfo{booktitle}{International Conference on Algorithmic
  Learning Theory}\/} (pp. \bibinfo{pages}{63--77}).
\newblock \bibinfo{organization}{Springer}.
\bibitem[{Gretton et~al.(2005{\natexlab{b}})Gretton, Smola, Bousquet, Herbrich,
  Belitski, Augath, Murayama, Pauls, Sch{\"o}lkopf \&
  Logothetis}]{gretton2005kernel}
\bibinfo{author}{Gretton, A.}, \bibinfo{author}{Smola, A.~J.},
  \bibinfo{author}{Bousquet, O.}, \bibinfo{author}{Herbrich, R.},
  \bibinfo{author}{Belitski, A.}, \bibinfo{author}{Augath, M.},
  \bibinfo{author}{Murayama, Y.}, \bibinfo{author}{Pauls, J.},
  \bibinfo{author}{Sch{\"o}lkopf, B.}, \& \bibinfo{author}{Logothetis, N.~K.}
  (\bibinfo{year}{2005}{\natexlab{b}}).
\newblock \bibinfo{title}{Kernel constrained covariance for dependence
  measurement.}
\newblock In {\it \bibinfo{booktitle}{AISTATS}\/} (pp.
  \bibinfo{pages}{112--119}).
\newblock volume~\bibinfo{volume}{10}.
\bibitem[{Hartigan(1990)}]{hartigan1990partition}
\bibinfo{author}{Hartigan, J.~A.} (\bibinfo{year}{1990}).
\newblock \bibinfo{title}{Partition models}.
\newblock {\it \bibinfo{journal}{Communications in Statistics Theory and
  Methods}\/},  {\it \bibinfo{volume}{19}\/}, \bibinfo{pages}{2745--2756}.
\bibitem[{Heckman(1976)}]{Heckman_causal_1976}
\bibinfo{author}{Heckman, J.~J.} (\bibinfo{year}{1976}).
\newblock \bibinfo{title}{The common structure of statistical models of
  truncation, sample selection and limited dependent variables and a simple
  estimator for such models}.
\newblock In {\it \bibinfo{booktitle}{Annals of economic and social
  measurement, volume 5, number 4}\/} (pp. \bibinfo{pages}{475--492}).
\newblock \bibinfo{publisher}{NBER}.
\bibitem[{Hoyer et~al.(2009)Hoyer, Janzing, Mooij, Peters \&
  Sch{\"o}lkopf}]{hoyer2009nonlinear}
\bibinfo{author}{Hoyer, P.~O.}, \bibinfo{author}{Janzing, D.},
  \bibinfo{author}{Mooij, J.~M.}, \bibinfo{author}{Peters, J.}, \&
  \bibinfo{author}{Sch{\"o}lkopf, B.} (\bibinfo{year}{2009}).
\newblock \bibinfo{title}{Nonlinear causal discovery with additive noise
  models}.
\newblock In {\it \bibinfo{booktitle}{Advances in Neural Information Processing
  Systems}\/} (pp. \bibinfo{pages}{689--696}).
\bibitem[{Hu et~al.(2018)Hu, Chen, Partovi~Nia, Chan \&
  Geng}]{Hu_CausalNips_2018}
\bibinfo{author}{Hu, S.}, \bibinfo{author}{Chen, Z.},
  \bibinfo{author}{Partovi~Nia, V.}, \bibinfo{author}{Chan, L.}, \&
  \bibinfo{author}{Geng, Y.} (\bibinfo{year}{2018}).
\newblock \bibinfo{title}{Causal inference and mechanism clustering of a
  mixture of additive noise models}.
\newblock In \bibinfo{editor}{S.~Bengio}, \bibinfo{editor}{H.~Wallach},
  \bibinfo{editor}{H.~Larochelle}, \bibinfo{editor}{K.~Grauman},
  \bibinfo{editor}{N.~Cesa-Bianchi}, \& \bibinfo{editor}{R.~Garnett} (Eds.),
  {\it \bibinfo{booktitle}{Advances in Neural Information Processing Systems
  31}\/} (pp. \bibinfo{pages}{5206--5216}).
\bibitem[{Janzing \& Scholkopf(2010)}]{janzing2010causal}
\bibinfo{author}{Janzing, D.}, \& \bibinfo{author}{Scholkopf, B.}
  (\bibinfo{year}{2010}).
\newblock \bibinfo{title}{Causal inference using the algorithmic markov
  condition}.
\newblock {\it \bibinfo{journal}{IEEE Transactions on Information Theory}\/},
  {\it \bibinfo{volume}{56}\/}, \bibinfo{pages}{5168--5194}.
\bibitem[{Kayikci \& Stix(2014)}]{kayikci2014causal}
\bibinfo{author}{Kayikci, Y.}, \& \bibinfo{author}{Stix, V.}
  (\bibinfo{year}{2014}).
\newblock \bibinfo{title}{Causal mechanism in transport collaboration}.
\newblock {\it \bibinfo{journal}{Expert systems with applications}\/},  {\it
  \bibinfo{volume}{41}\/}, \bibinfo{pages}{1561--1575}.
\bibitem[{Lawrence(2005)}]{lawrence2005probabilistic}
\bibinfo{author}{Lawrence, N.} (\bibinfo{year}{2005}).
\newblock \bibinfo{title}{Probabilistic non-linear principal component analysis
  with gaussian process latent variable models}.
\newblock {\it \bibinfo{journal}{Journal of Machine Learning Research}\/},
  {\it \bibinfo{volume}{6}\/}, \bibinfo{pages}{1783--1816}.
\bibitem[{Li et~al.(2019)Li, Mccormick \& Clark}]{Li_BGLASSO_2019}
\bibinfo{author}{Li, Z.}, \bibinfo{author}{Mccormick, T.}, \&
  \bibinfo{author}{Clark, S.} (\bibinfo{year}{2019}).
\newblock \bibinfo{title}{Bayesian joint spike-and-slab graphical lasso}.
\newblock In {\it \bibinfo{booktitle}{International Conference on Machine
  Learning}\/} (pp. \bibinfo{pages}{3877--3885}).
\newblock \bibinfo{organization}{PMLR}.
\bibitem[{Liu \& Chan(2016)}]{liu2016causal}
\bibinfo{author}{Liu, F.}, \& \bibinfo{author}{Chan, L.}
  (\bibinfo{year}{2016}).
\newblock \bibinfo{title}{Causal discovery on discrete data with extensions to
  mixture model}.
\newblock {\it \bibinfo{journal}{ACM Transactions on Intelligent Systems and
  Technology (TIST)}\/},  {\it \bibinfo{volume}{7}\/}, \bibinfo{pages}{21}.
\bibitem[{Mooij et~al.(2016)Mooij, Peters, Janzing, Zscheischler \&
  Sch{\"o}lkopf}]{mooij2016distinguishing}
\bibinfo{author}{Mooij, J.~M.}, \bibinfo{author}{Peters, J.},
  \bibinfo{author}{Janzing, D.}, \bibinfo{author}{Zscheischler, J.}, \&
  \bibinfo{author}{Sch{\"o}lkopf, B.} (\bibinfo{year}{2016}).
\newblock \bibinfo{title}{Distinguishing cause from effect using observational
  data: methods and benchmarks}.
\newblock {\it \bibinfo{journal}{The Journal of Machine Learning Research}\/},
  {\it \bibinfo{volume}{17}\/}, \bibinfo{pages}{1103--1204}.
\bibitem[{Murray et~al.(2011)Murray, Lopez, Black, Ahuja, Ali, Baqui, Dandona,
  Dantzer, Das, Dhingra et~al.}]{Murray_VerbalAutopsyData_2011}
\bibinfo{author}{Murray, C.~J.}, \bibinfo{author}{Lopez, A.~D.},
  \bibinfo{author}{Black, R.}, \bibinfo{author}{Ahuja, R.},
  \bibinfo{author}{Ali, S.~M.}, \bibinfo{author}{Baqui, A.},
  \bibinfo{author}{Dandona, L.}, \bibinfo{author}{Dantzer, E.},
  \bibinfo{author}{Das, V.}, \bibinfo{author}{Dhingra, U.} et~al.
  (\bibinfo{year}{2011}).
\newblock \bibinfo{title}{Population health metrics research consortium gold
  standard verbal autopsy validation study: design, implementation, and
  development of analysis datasets}.
\newblock {\it \bibinfo{journal}{Population health metrics}\/},  {\it
  \bibinfo{volume}{9}\/}, \bibinfo{pages}{1--15}.
\bibitem[{Neyman(1923)}]{Neyman_Causal_1923}
\bibinfo{author}{Neyman, J.~S.} (\bibinfo{year}{1923}).
\newblock \bibinfo{title}{On the application of probability theory to
  agricultural experiments.}
\newblock {\it \bibinfo{journal}{Statistical Science}\/},  {\it
  \bibinfo{volume}{5}\/}, \bibinfo{pages}{465--472}.
\bibitem[{Pearl(1986)}]{pearl1986fusion}
\bibinfo{author}{Pearl, J.} (\bibinfo{year}{1986}).
\newblock \bibinfo{title}{Fusion, propagation, and structuring in belief
  networks}.
\newblock {\it \bibinfo{journal}{Artificial intelligence}\/},  {\it
  \bibinfo{volume}{29}\/}, \bibinfo{pages}{241--288}.
\bibitem[{Pearl(2009)}]{pearl2009causality}
\bibinfo{author}{Pearl, J.} (\bibinfo{year}{2009}).
\newblock {\it \bibinfo{title}{Causality}\/}.
\newblock \bibinfo{publisher}{Cambridge university press}.
\bibitem[{Pelleg et~al.(2000)Pelleg, Moore et~al.}]{pelleg2000x}
\bibinfo{author}{Pelleg, D.}, \bibinfo{author}{Moore, A.~W.} et~al.
  (\bibinfo{year}{2000}).
\newblock \bibinfo{title}{X-means: Extending k-means with efficient estimation
  of the number of clusters.}
\newblock In {\it \bibinfo{booktitle}{International Conference on Machine
  Learning}\/} (pp. \bibinfo{pages}{727--734}).
\newblock volume~\bibinfo{volume}{1}.
\bibitem[{R{\"u}cker \& Schumacher(2008)}]{rucker2008simpson}
\bibinfo{author}{R{\"u}cker, G.}, \& \bibinfo{author}{Schumacher, M.}
  (\bibinfo{year}{2008}).
\newblock \bibinfo{title}{Simpson's paradox visualized: the example of the
  rosiglitazone meta-analysis}.
\newblock {\it \bibinfo{journal}{BMC medical research methodology}\/},  {\it
  \bibinfo{volume}{8}\/}, \bibinfo{pages}{34}.
\bibitem[{Schadt et~al.(2005)Schadt, Lamb, Yang, Zhu, Edwards, GuhaThakurta,
  Sieberts, Monks, Reitman, Zhang et~al.}]{schadt2005integrative}
\bibinfo{author}{Schadt, E.~E.}, \bibinfo{author}{Lamb, J.},
  \bibinfo{author}{Yang, X.}, \bibinfo{author}{Zhu, J.},
  \bibinfo{author}{Edwards, S.}, \bibinfo{author}{GuhaThakurta, D.},
  \bibinfo{author}{Sieberts, S.~K.}, \bibinfo{author}{Monks, S.},
  \bibinfo{author}{Reitman, M.}, \bibinfo{author}{Zhang, C.} et~al.
  (\bibinfo{year}{2005}).
\newblock \bibinfo{title}{An integrative genomics approach to infer causal
  associations between gene expression and disease}.
\newblock {\it \bibinfo{journal}{Nature genetics}\/},  {\it
  \bibinfo{volume}{37}\/}, \bibinfo{pages}{710--717}.
\bibitem[{Sch{\"o}lkopf et~al.(2012)Sch{\"o}lkopf, Janzing, Peters, Sgouritsa,
  Zhang \& Mooij}]{scholkopf2012causal}
\bibinfo{author}{Sch{\"o}lkopf, B.}, \bibinfo{author}{Janzing, D.},
  \bibinfo{author}{Peters, J.}, \bibinfo{author}{Sgouritsa, E.},
  \bibinfo{author}{Zhang, K.}, \& \bibinfo{author}{Mooij, J.}
  (\bibinfo{year}{2012}).
\newblock \bibinfo{title}{On causal and anticausal learning}.
\newblock {\it \bibinfo{journal}{arXiv preprint arXiv:1206.6471}\/}, .
\bibitem[{Schwarz et~al.(1978)}]{schwarz1978estimating}
\bibinfo{author}{Schwarz, G.} et~al. (\bibinfo{year}{1978}).
\newblock \bibinfo{title}{Estimating the dimension of a model}.
\newblock {\it \bibinfo{journal}{The annals of statistics}\/},  {\it
  \bibinfo{volume}{6}\/}, \bibinfo{pages}{461--464}.
\bibitem[{Shimizu et~al.(2006)Shimizu, Hoyer, Hyv{\"a}rinen \&
  Kerminen}]{shimizu2006linear}
\bibinfo{author}{Shimizu, S.}, \bibinfo{author}{Hoyer, P.~O.},
  \bibinfo{author}{Hyv{\"a}rinen, A.}, \& \bibinfo{author}{Kerminen, A.}
  (\bibinfo{year}{2006}).
\newblock \bibinfo{title}{A linear non-gaussian acyclic model for causal
  discovery}.
\newblock {\it \bibinfo{journal}{Journal of Machine Learning Research}\/},
  {\it \bibinfo{volume}{7}\/}, \bibinfo{pages}{2003--2030}.
\bibitem[{Simpson(1951)}]{simpson1951interpretation}
\bibinfo{author}{Simpson, E.~H.} (\bibinfo{year}{1951}).
\newblock \bibinfo{title}{The interpretation of interaction in contingency
  tables}.
\newblock {\it \bibinfo{journal}{Journal of the Royal Statistical Society:
  Series B (Methodological)}\/},  {\it \bibinfo{volume}{13}\/},
  \bibinfo{pages}{238--241}.
\bibitem[{Wood et~al.(1993)Wood, Booth \& Butler}]{wood1993saddlepoint}
\bibinfo{author}{Wood, A.~T.}, \bibinfo{author}{Booth, J.~G.}, \&
  \bibinfo{author}{Butler, R.~W.} (\bibinfo{year}{1993}).
\newblock \bibinfo{title}{Saddlepoint approximations to the cdf of some
  statistics with nonnormal limit distributions}.
\newblock {\it \bibinfo{journal}{Journal of the American Statistical
  Association}\/},  {\it \bibinfo{volume}{88}\/}, \bibinfo{pages}{680--686}.
\bibitem[{Wu(2010)}]{wu2010linking}
\bibinfo{author}{Wu, W.~W.} (\bibinfo{year}{2010}).
\newblock \bibinfo{title}{Linking bayesian networks and pls path modeling for
  causal analysis}.
\newblock {\it \bibinfo{journal}{Expert Systems with Applications}\/},  {\it
  \bibinfo{volume}{37}\/}, \bibinfo{pages}{134--139}.
\bibitem[{Zhang \& Hyv{\"a}rinen(2009)}]{zhang2009identifiability}
\bibinfo{author}{Zhang, K.}, \& \bibinfo{author}{Hyv{\"a}rinen, A.}
  (\bibinfo{year}{2009}).
\newblock \bibinfo{title}{On the identifiability of the post-nonlinear causal
  model}.
\newblock In {\it \bibinfo{booktitle}{Conference on Uncertainty in Artificial
  Intelligence}\/} (pp. \bibinfo{pages}{647--655}).
\newblock \bibinfo{organization}{AUAI Press}.

\end{thebibliography}

\newpage 

\begin{center}
\Large
    Appendix
\end{center}

\textbf{Proof of Theorem~\ref{theo:conv}}\\
The proof is a  multi-stage Gibbs sampler adaptation for the clustering case with varying cluster components. 
First we ensure that sampling from the discrete multivariate posterior $p(\z\mid\mmu, \ttheta, k)$ and continuous multivariate  $p(\mmu\mid\z,\ttheta, k)$ converges to the joint $p(\mmu,\z\mid\ttheta, k)$. 
Note that $p(\mmu\mid\ttheta, \z,k)$ is multivariate Gaussian and $p(\z\mid\ttheta, \mmu,k)$ is discrete with support $\{1,\ldots,k\}^N$.
Define the positive Markov transition kernel 
$$k(\z \mid \z') = \int \cdots \int p(\z\mid \mmu, \ttheta,k) p(\mmu\mid \z',\ttheta,k) d\mmu.$$ This transition kernel is equivalent to taking intermediate samples from $\mmu_t\sim p(\mmu\mid\z_{t-1},\ttheta,k)$ at iteration $t$ and drawing $\z_t\sim p(\z\mid\mmu_{t},\ttheta,k)$.

It is easy to check that $k(\z\mid\z')$ is reversible and hence  invariant with respect to the marginal $p(\z \mid\ttheta,k)$.

Now suppose we sample from the multivariate $p(\z\mid\mmu,\ttheta,k)$ using univariate multinomial samplers. Let $k_1(\z\mid\z')$ be the transition kernel of a univariate Gibbs sampler of $p(\z\mid\mmu,\ttheta,k)$ in increasing order $z_1,\ldots, z_N$, i.e.
$$ k_1(\z\mid\z') =  p(z_1'\mid z_2, \ldots, z_N,\mmu,\ttheta,k) p(z'_2,\mid z'_1,z_3,\ldots, z_N,\mmu,\ttheta,k )\cdots p(z'_N\mid z'_1, \ldots, z'_{N-1},\mmu,\ttheta,k)   
 $$
\begin{eqnarray*}
\sum_{z_1,\ldots, z_n} &&k_1(\z\mid\z') p(\z\mid\mmu,\ttheta,k)\\  
= \sum_{z_1,\ldots, z_n} && p(z_1'\mid z_2, \ldots, z_N,\mmu,\ttheta,k) \cdots p(z'_N\mid z'_1, \ldots, z'_{N-1},\mmu,\ttheta,k)\\
&& p(z_1\mid z_2,\ldots, z_N,\mmu,\ttheta,k) p(z_2,\ldots, z_N\mid\mmu,\ttheta,k)\\
= \sum_{z_2,\ldots, z_n} && p(z'_2\mid z'_1, \ldots, z_N,\mmu,\ttheta,k) p(z'_n\mid z'_1, \ldots, z'_{N-1},\mmu,\ttheta,k) p(z'_1,z_2,\ldots, z_N)
 \end{eqnarray*}
in which we integrated over $z_1$ and decomposed 
$$p(z'_1, z_2,\ldots, z_N\mid \mmu,\ttheta,k) = p(z'_1\mid z_2, \ldots, z_N,\mmu,\ttheta,k) p(z_2,\ldots, z_N\mid\mmu,\ttheta).$$
Continue by decomposing 
$$p(z'_1, z_2,\ldots, z_N\mid \mmu,\ttheta,k) = p(z_2\mid z'_1,z_3,\ldots, z_N \mid \mmu,\ttheta,k) p(z'_1,z_3,\ldots, z_N \mid \mmu,\ttheta,k)$$
and summing over $z_2$. Repeating this re-arranging and summing over $z_3, \ldots,z_N$ ends up with
$p(z'_1,\ldots, z'_N\mid \mmu,\ttheta,k) = p(\z'\mid\mmu,\ttheta,k)$.

A similar argument applies to $\mmu_t \sim p(\mmu\mid\z,\ttheta)$ to replace the intermediate sampler with univariate conditional samplers and to show
$$p(\mmu'\mid \z, \ttheta) = \int \cdots \int k_2(\mmu\mid\mmu') p(\mmu\mid\z,\ttheta)d\mmu$$
in which 
$$k_2(\mmu\mid\mmu')=p(\mu_1'\mid \mu_2,\ldots,\mu_c,\z,\ttheta)p(\mu'_2\mid\mu'_1,\mu_3,\ldots,\mu_c,\z,\ttheta)\cdots p(\mu'_c\mid \mu'_1,\ldots,\mu'_{c-1}).$$
Simple posterior calculations show these univariate distributions that construct the kernel $k_1$  are multinomial  $z_i\sim \M(1,\ppi_i)$ with probabilities 
$$\ppi_i = \left[{\phi\left({\theta_i-\mu_1\over \sigma}\right) \over \sum_{c=1}^k \phi\left({\theta_i-\mu_c\over \sigma}\right) },\ldots, {\phi\left({\theta_i-\mu_c\over \sigma}\right) \over \sum_{c=1}^k \phi\left({\theta_i-\mu_c\over \sigma}\right)} \right],$$
and the intermediate univariate samplers are Gaussian with  mean $\bar\theta_c (1 + {1\over N_c\kappa\tau^2})^{-1}$ and variance $\sigma^2(N_c+{1\over \kappa\tau^2})^{-1}$. 
The label update step has $k_1$ and the mean update has $k_2$ transition kernel. Implementing these two steps  sequentially is equivalent to a chain with the composition transition kernel $k_1 \circ k_2$. The last step is to margin over the cluster components $k$.

Assume  a discrete uniform prior on $k\in\{K-\Delta, K+\Delta\}$ which allows to define a posterior proportional to the likelihood. Note that $k$ does not affect the dimension of the marginalized posterior $p(\z\mid\ttheta, k)$, otherwise trans-dimensional samplers need to be developed. Marginalizing over $k$ adds another step to the algorithm and implies sampling $k\sim\M(1,\ppi)$ in which  
$$ \ppi = \left[ {\exp\{ \ell(\z\mid\ttheta, k=K-\Delta)\} \over \sum\limits_{k=K-\Delta}^{K+\Delta}\exp\{\ell(\z\mid\ttheta, k)\}} ,\ldots, {\exp\{ \ell(\z\mid\ttheta, k=K-\Delta)\} \over \sum\limits_{k=K-\Delta}^{K+\Delta}\exp\{\ell(\z\mid\ttheta, k)\}} \right]. ~~\blacksquare$$

\textbf{Proof of Theorem~\ref{theo:HSIC}}\\ The product partition model \eqref{eq:prodpart} imposes mutually independent random pairs $(x_i, y_i)$ across clusters. Therefore, given a certain causal direction their projections $\theta_i = g(x_i, y_i)$ are independent across clusters too. A similar argument holds for $\HSIC$ as a function of $\theta_i$. This allows us to use the asymptotic results of the homogeneous case \citep{gretton2005measuring} in each cluster $c$ and combine them using a  product partition independence assumption.

For large $n_c$ inside cluster $c$ 
$$n_c\HSIC_c \sim \sum_{l=1}^\infty \lambda_l z_l^2 \approx \Gamma(\alpha, \beta)$$ in which $\lambda_l$ are constants and $z_l$ are independent standard Gaussian random variables with $\mu_c = \E(\HSIC_c), \sigma^2_c = \V(\HSIC_c)$. Define the aggregated test statistic 
$$ t = \sum_{c=1}^k n_c\HSIC_c = \sum_{c=1}^k \sum_{l=1}^\infty \lambda_{cl} z_{cl}^2 $$
which is clearly another countable sum  $\sum\limits_{m=1}^\infty  \gamma_m w^2_m$ after swapping the sum order and re-arranging terms  $m = (l-1)k+c. ~~ \blacksquare$

\end{document}